\documentclass[twocolumn,aps,amsmath,amssymb,color,superscriptaddress,prx,footinbib,longbibliography]{revtex4-2}
\usepackage{stix}
\usepackage[colorlinks,citecolor=blue,linkcolor=blue,urlcolor=blue]{hyperref}
\usepackage{graphicx}
\usepackage{dcolumn}
\usepackage{multirow}
\usepackage{bm}
\usepackage{times}
\usepackage{tabularx}
\usepackage{footmisc}
\usepackage{xcolor}
\usepackage{float}
\usepackage{tabularx}
\usepackage{url}
\usepackage{hyperref}
\usepackage{ulem}
\usepackage{titlesec}
\usepackage[english]{babel}

\begin{document}

\title{Fractional Chern insulator states in an isolated flat band of zero Chern number}

\author{Zuzhang Lin}
\thanks{These authors contribute equally to the work.}
\affiliation{New Cornerstone Science Lab, Department of Physics, The University of Hong Kong, Hong Kong, China}
\affiliation{HK Institute of Quantum Science \& Technology, The University of Hong Kong, Hong Kong, China}

\author{Hongyu Lu}
\thanks{These authors contribute equally to the work.}
\affiliation{New Cornerstone Science Lab, Department of Physics, The University of Hong Kong, Hong Kong, China}
\affiliation{HK Institute of Quantum Science \& Technology, The University of Hong Kong, Hong Kong, China}

\author{Wenqi Yang}
\affiliation{New Cornerstone Science Lab, Department of Physics, The University of Hong Kong, Hong Kong, China}
\affiliation{HK Institute of Quantum Science \& Technology, The University of Hong Kong, Hong Kong, China}

\author{Dawei Zhai}
\affiliation{New Cornerstone Science Lab, Department of Physics, The University of Hong Kong, Hong Kong, China}
\affiliation{HK Institute of Quantum Science \& Technology, The University of Hong Kong, Hong Kong, China}

\author{Wang Yao}
\email{wangyao@hku.hk}
\affiliation{New Cornerstone Science Lab, Department of Physics, The University of Hong Kong, Hong Kong, China}
\affiliation{HK Institute of Quantum Science \& Technology, The University of Hong Kong, Hong Kong, China}

\date{\today}
\begin{abstract}
A flat band with Chern number $C=0$, and well isolated from the rest of Hilbert space by a gap much larger than interaction strength, is a context that has not been regarded as relevant for fractional quantum Hall physics.
In this work, we demonstrate the emergence of the fractional Chern insulator (FCI) states in such a trivial flat band, using large-scale exact diagonalization (ED) and infinite density matrix renormalization group (iDMRG) simulations.
The $C=0$ isolated flat band is hosted by an anisotropic fluxed dice lattice, which is reducible to a two-orbital honeycomb lattice with up to third nearest-neighbor hopping.
Both the quantum metric and Berry curvature of the $C=0$ flat band have a sharp peak at the $\Gamma$ point, whereas in the rest of the Brillouin zone (BZ) they mimic the quantum geometry of the lowest Landau level.
We consider nearest-neighbor repulsion that is weak enough to ensure the isolated-band limit is always satisfied, where the Hartree-Fock  band renormalization leaves the quantum geometry of the flat band unchanged.
From the projected ED simulations at $\nu_\mathrm{F}=2/3$ electron filling of the flat band (i.e. $1/3$ hole filling), we find the unexpected FCI with 3-fold ground-state degeneracy and $\sigma_\mathrm{H}=-1/3 (e^2/h)$.
The momentum space carrier distribution shows that the quantum metric peak tends to push the interacting holes away from $\Gamma$ point towards the BZ regions with the nearly ``ideal'' quantum geometry, underlying the formation of FCI in the $C=0$ flat band.
In comparison, we find only trivial ground states at $\nu_\mathrm{F}=1/3$, when the $\Gamma$ peak becomes less avoidable.
Besides, when tuning the single-particle anisotropy such that the quantum geometry of the $C=0$ flat band becomes less sharp around $\Gamma$, we find the ground state becomes a charge density wave with tripled unit cell at $\nu_\mathrm{F}=2/3$.
Our two-band iDMRG simulations further corroborate the  FCI in the isolated $C=0$ flat band, demonstrating in such parameter regime the fractionally quantized charge pumping upon flux insertion as well as the momentum-resolved entanglement spectrum characteristic of the $1/3$ Laughlin state.
Our work not only shows that FCI states could exist without the single-particle topology, relying on the more intrinsic quantum geometry, but also suggests a possibly generic scenario to explore FCIs in $C=0$ isolated flat band.
\end{abstract}
\maketitle

\section{Introduction}

The topological equivalence of Chern bands in crystals to Landau levels in two dimensional electron gas \cite{thouless_quantized_1982},
characterized by their common Chern number, underlies the realization of integer quantum anomalous Hall effect at zero magnetic field \cite{haldane_model_1988,chang_experimental_2013}.
In the isolated flat band limit (i.e. bandgaps $\gg$ interaction $\gg$ bandwidth, mimicking the energy landscape of a Landau level), Chern band naturally provides a platform to explore the fractional quantum anomalous Hall (FQAH) effect \cite{tang_high-temperature_2011,wu_zoology_2012,sun_nearly_2011,neupert_fractional_2011,sheng_fractional_2011,wang_fractional_2011,regnault_fractional_2011,xiao_interface_2011},
also termed as fractional Chern insulator (FCI) \cite{regnault_fractional_2011, BERGHOLTZ2013_FCI_review}.
Apart from band topology, the quantum geometry in the projected Hilbert space is known to play a crucial role.
Mimicking the momentum-space quantum geometry of Landau level, including uniform Berry curvature, the trace condition \cite{roy_band_2014}, 
GMP algebra \cite{girvin_magneto-roton_1986,parameswaran_fractional_2012},
has served as a working guideline to screen Chern bands for FCI \cite{ledwith_fractional_2020,jackson_geometric_2015,claassen_position-momentum_2015,lee_band_2017,mera_engineering_2021,mera_kahler_2021,ozawa_relations_2021,mera_relating_2022,zhang_revealing_2022,varjas_topological_2022, wang_exact_2021,claassen_position-momentum_2015},
for which vortexability as another characteristic of Landau level further provides a unifying perspective in real-space \cite{ledwith_vortexability_2023-1}.
It is worth noting that a time-reversal pair of flat Chern bands can host fractional topological insulators that support fractionalized excitation in presence of time-reversal symmetry \cite{levin2009fractional,neupert2011fractional,kang2024evidence}, whereas spontaneous time-reversal symmetry breaking in such states can lead to FCI \cite{neupert2011fractional}.
FCI has also been explored upon band inversions in a multi-band manifold \cite{simon_fractional_2015,hu_fractional_2018,yang_fractional_2024}, which is the opposite limit to the isolated flat band scenario.
Through introducing effective lattice hopping \cite{simon_fractional_2015}
or spontaneous charge ordering \cite{kourtis_symmetry_2018},
interaction can drive the topological transitions to develop a Chern band in the first place, which can sustain FCI at its partial filling.
It has also been proposed that higher angular momentum band inversions can lead to fractional quantum Hall states at integer band filling \cite{hu_fractional_2018}.
Backed up by these extensive theoretical efforts, as well as first-principle-based investigations of realizations in moir{\'e} materials \cite{li_spontaneous_2021,wu_topological_2019,ledwith_fractional_2020,abouelkomsan2020particle,zhang2019nearly},
the groundbreaking observation of FCI at zero magnetic field was first achieved in twisted bilayer MoTe$_2$~\cite{cai_signatures_2023,zeng_thermodynamic_2023,park_observation_2023,xu_observation_2023},
and subsequently in graphene/hBN moir\'e \cite{lu_fractional_2024,lu_extended_2025}.
FCI states stabilized by finite magnetic field have also been observed earlier in graphene moir\'e systems \cite{spanton2018observation,xie2021fractional}.
The experiments suggest that FCI in these systems states emerge from partially filled Chern bands.

In this work, we present a comprehensive study to demonstrate the evidence of FCI in an isolated flat band of zero Chern number, a context that has not been considered relevant for fractional quantum Hall physics.
The $C=0$ isolated flat band is hosted by an anisotropic fluxed dice lattice, with the hopping anisotropy characterized by the factor $\eta \in [0,1]$ multiplied on the hopping
amplitudes along one of the three directions, where $\eta=0$ corresponds to the limit of disconnected chains.
For $\eta < 1$, all three bands have zero Chern number, and the middle one has exact flatness and is separated from the adjacent lower dispersive band by a gap $\Delta \propto (1-\eta)^2$. We focus on the regime where the top most dispersive band is far detuned and virtually irrelevant, such that the dice lattice can reduce to a two-band honeycomb lattice with up to third nearest-neighbor hopping.
The quantum geometry of the isolated flat band has the following feature:   quantum metric and Berry curvature both have a sharp peak at the $\Gamma$ point, while in the rest of the BZ the quantum geometry mimics that of the lowest Landau level.

We consider nearest-neighbor repulsion that is weak enough to ensure the isolated-band limit is always satisfied, where the Hartree-Fock band renormalization leaves the quantum geometry of the flat band unchanged and only introduces a tiny band width with the energy minimum at $\Gamma$.
From the projected ED simulations at $\nu_\mathrm{F}=2/3$ electron filling of the $C=0$ flat band, we find the unexpected FCI with 3-fold ground-state degeneracy and $\sigma_\mathrm{H}=-1/3(e^2/h)$.
The momentum space carrier distribution in the FCI ground states suggests that it is the holes that are preferentially occupying the BZ region (away from $\Gamma$) with the nearly ideal quantum geometry.
Since the particle-hole transformation of the projected Hamiltonian will introduce an effective hole dispersion that exactly cancels the Hartree-Fock dispersion, the holes are actually seeing an exactly-flat $C=0$ band where the
quantum metric peak tends to push the interacting holes away from $\Gamma$ point.
Our ED calculation of this hole scenario at the equivalent $\nu_\mathrm{F}^\ast=1/3$ hole filling obtain the same FCI spectra.
In essence, the FCI comes from the coordination of the quantum metric and Berry curvature in the $C=0$ band.
In comparison, we find only trivial ground states at $\nu_\mathrm{F}=1/3$ ($\nu_\mathrm{F}^\ast=2/3$), when the $\Gamma$ peak of Berry curvature becomes less avoidable.
Upon tuning the single-particle anisotropy such that the quantum geometry of the $C=0$ flat band becomes less sharp around $\Gamma$, we find that the FCI ground state at $\nu_\mathrm{F}=2/3$ undergoes a transition to a charge density wave (CDW) with tripled unit cell.

Our two-band infinite density matrix renormalization group (iDMRG) results further support this exotic FCI at the isolated $C=0$ flat band limit, demonstrating in such parameter regime the fractionally quantized charge pumping upon adiabatic flux insertion as well as the momentum-resolved entanglement spectrum showing the characteristic edge counting $\{ 1,1,2,3,5... \}$ of the $1/3$ Laughlin state.
Our work not only demonstrates that the FCI could exist without the single-particle topology and rely on the more intrinsic quantum geometry, but also points to a possibly generic scenario to explore FCIs in isolated flat $C=0$ band.

\section{Isolated flat band of zero Chern number in anisotropic lattice models}
\label{sec_model}

\subsection{Fluxed dice lattice with anisotropic hopping}

\begin{figure}
\centering
\includegraphics[width=1\columnwidth]{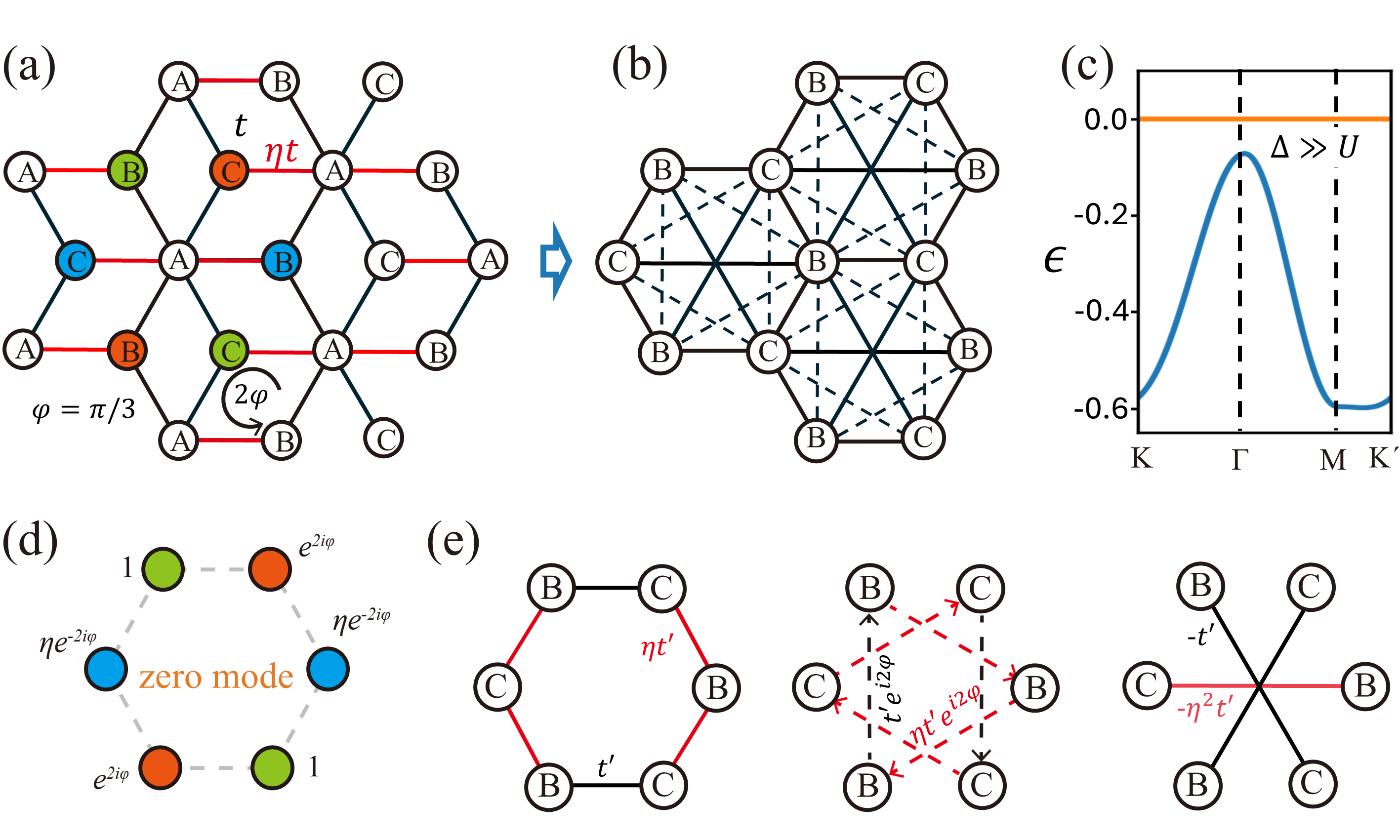}
\caption{ \textbf{Lattice models and band structures.}
(a) Schematic illustration of the dice lattice with anisotropic hopping, with hopping along $x$ direction set to $\eta t$, otherwise $t$. (b) The effective two-orbital model where the high-energy A orbitals are perturbatively eliminated. (c) Band structures at $\eta=0.4$ as an example, with both bands having $C=0$. (d) Zero modes that span the isolated flat band. (e) Schematic diagram of hoppings among nearest, next nearest, and next-next nearest neighbor sites of the effective two-orbital model in (b). $t^{\prime}=-t^2/E_A$.
The isolated band limit will be considered, where interactions - being much smaller than the gap - leave the quantum geometry and trivial topology of the flat band unchanged.
}
\label{Fig_band}
\end{figure}

We start from a three-orbital tight-binding model with nearest-neighbor hopping only on a dice lattice geometry \cite{vidal1998aharonov,sutherland1986localization} [Fig. \ref{Fig_band}(a)]. The model Hamiltonian reads
\begin{eqnarray}\label{Eq_H0}
\hat{H}_0 =&& \sum_l \epsilon_A\,\hat{A}^{\dagger}_{l} \hat{A}_{l} \\
            &&- \sum_{\langle l,m \rangle} \left( t_{\langle l,m \rangle} e^{i \phi_1^{l,m}} \hat{A}^{\dagger}_l \hat{B}_m
+ t_{\langle l,m \rangle}\, e^{i \phi_2^{l,m}} \hat{A}^{\dagger}_l \hat{C}_m + h.c. \right). \notag
\end{eqnarray}
Here, the B and C orbitals are degenerate, and we set their onsite energy to zero. $\epsilon_A >0 $ is the A orbital detuning, which is taken to be much larger than all other energy scales, so that this orbital is virtually unoccupied.
$\phi^{\langle l,m \rangle}$ are the phases of hoppping matrix elements determined by a $2\pi/3$ flux threading each rhombus [Fig. \ref{Fig_band}(a)].
An anisotropy is introduced on the hopping strength: the hopping matrix elements along $x$ direction are multiplied with a dimensionless factor $\eta$, i.e., $t_{\langle l,m \rangle} = \eta t$ [c.f. red bonds in Fig. \ref{Fig_band}(a)], while hopping strength along the other two directions remains as $t$.

The momentum space  Hamiltonian of such a dice lattice geometry reads
\begin{equation}
\hat{H}_0(\mathbf{k})=\left(\begin{array}{ccc}
\epsilon_A & f(\mathbf{k}) & g(\mathbf{k}) \\
f^*(\mathbf{k}) & 0 & 0 \\
g^*(\mathbf{k}) & 0 & 0
\end{array}\right),
\end{equation}
where
\begin{equation}
\begin{aligned}
& f(\mathbf{k})=-t\left[\eta e^{i \mathbf{k} \cdot \mathbf{d}_1}+e^{i 2 \varphi} e^{i \mathbf{k} \cdot \mathbf{d}_3}+e^{-i 2 \varphi} e^{i \mathbf{k} \cdot \mathbf{d}_5}\right]  \\
& g(\mathbf{k})=-t\left[e^{i \varphi}e^{i \mathbf{k} \cdot \mathbf{d}_2}+\eta e^{-i3 \varphi} e^{i \mathbf{k} \cdot \mathbf{d}_4}+e^{-i \varphi} e^{i \mathbf{k} \cdot \mathbf{d}_6}\right].
\end{aligned}
\end{equation}
Here $\varphi=\pi/3$, $\mathbf{d}_1=a(1,0)$ with $a$ being the length of the nearest-neighbor bond and $\mathbf{d}_2$ to $\mathbf{d}_5$ obtained by successive 60-degree counterclockwise rotations of $\mathbf{d}_1$.
The eigenenergies in ascending order are $\epsilon_1=[\epsilon_A-\sqrt{\epsilon_A^2+4(|f(\mathbf{k})|^2+|g(\mathbf{k})|^2}]/2$, $\epsilon_2=0$ and $\epsilon_3=[\epsilon_A+\sqrt{\epsilon_A^2+4(|f(\mathbf{k})|^2+|g(\mathbf{k})|^2}]/2$.
The top band (not plotted) is detuned by $\sim \epsilon_A$ and can be well neglected.
The middle flat band of the three-orbital model consists of zero modes of the Hamiltonian [Fig. \ref{Fig_band}(d)]. It is separated from the lower dispersive band by a gap with the minimum at $\Gamma$ point:
\begin{align}\label{Eq_gap}
    \Delta&=(\sqrt{\epsilon_A^2+8t^2(\eta-1)^2}-\epsilon_A)/2\approx2(\eta-1)^2t^2/\epsilon_A.
\end{align}

\begin{figure}
\centering
\includegraphics[width=1\columnwidth]{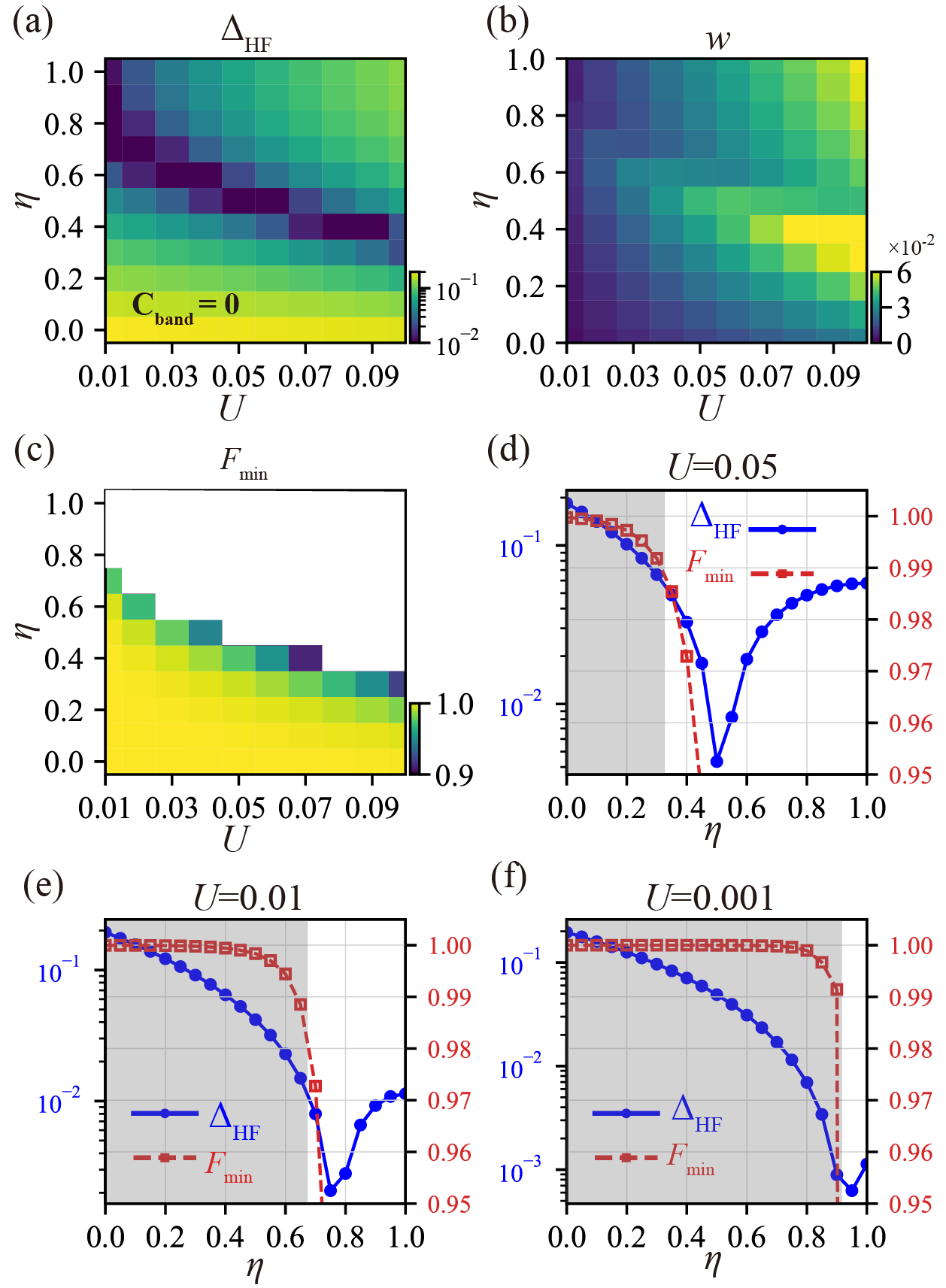}
\caption{\textbf{Hartree-Fock results at $\nu=1$} (filled lower dispersive band and empty flat band).
(a) Hartree-Fock band gap $\Delta_{\rm HF}$ as a function of interaction strength $U$ and anisotropy parameter $\eta$. (b) Band width $w$.
(c) Minimum Bloch function fidelity $F_{\rm min} = \min_{\rm BZ} F(\bm k)$, where $F(\bm k) \equiv | \langle \psi_{\rm HF} (\bm k) | \psi_{\rm flat} (\bm k) \rangle | $, with $\psi_{\rm HF}$  the Hartree-Fock wavefunction of the upper flat band and $\psi_{\rm flat}$ the Bloch function in the non-interacting limit as given in Eq.~(\ref{Eq_wave}). (d)-(e) Dependence of the Hartree-Fock band gap and minimum fidelity on $\eta$ for $U=0.05$ (d), $U=0.01$ (e), and $U=0.001$ (f).
The gray shaded area denotes the isolated band limit with $F_{\rm min} \geq 0.99$.
}\label{Fig_HF}
\end{figure}

\begin{figure*}
\centering
\includegraphics[width=1\textwidth]{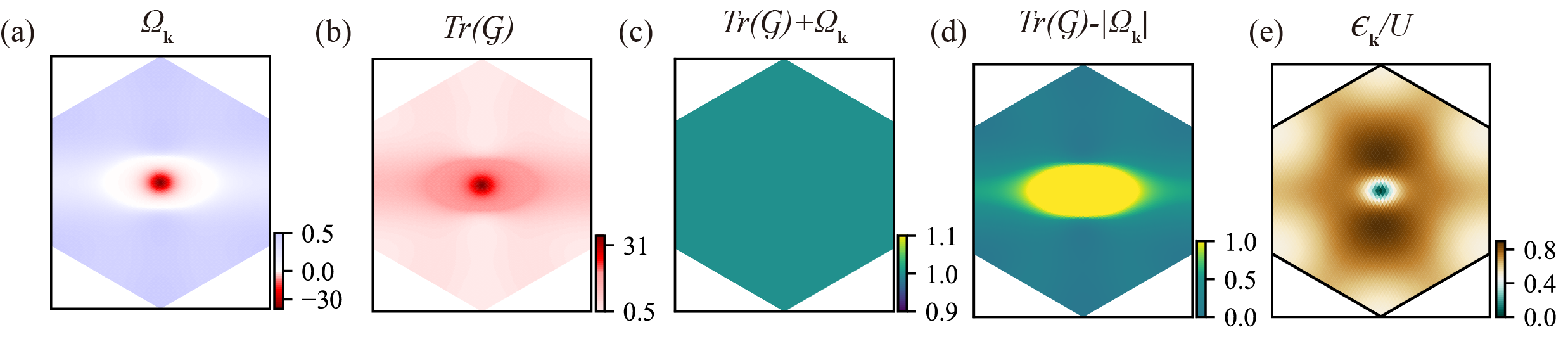}
\caption{ \textbf{Propoerties of isolated $C=0$ flat band at $\eta=0.7$}. (a) Berry curvature $\Omega_{\mathbf{k}}$, (b) Trace of quantum metric tensor $\mathcal{G}$, (c) ${\rm Tr} \mathcal{G}+\Omega_{\mathbf{k}}$, (d) ${\rm Tr} \mathcal{G}-|\Omega_{\mathbf{k}}|$, and (e) Hartree Fock renormalized dispersion relation $\epsilon_\mathbf{k}$,
under an interaction strength $U=0.001$. Such a small $U$, to be adopted in ED calculations, ensures the isolated band limit where $\mathcal{G}$ and $\Omega_{\mathbf{k}}$ remain the same as in non-interacting case.
${\rm Tr} \mathcal{G}$ and $\Omega_{\mathbf{k}}$ are in unit of $a^2$.
}\label{Fig_BC}
\end{figure*}

The three-orbital model here is a variant of the tight-binding model initially proposed for describing the lowest minibands of the twisted bilayer MoTe$_2$ moir\'e superlattice \cite{yu_giant_2020}.
With nearest-neighbor hoppings between all three orbitals, this model can well capture the Chern numbers and dispersions of the three lowest minibands in twisted MoTe$_2$ \cite{wu_topological_2019}, underlying the discovery of FCI in twisted MoTe$_2$~\cite{cai_signatures_2023,zeng_thermodynamic_2023,park_observation_2023,xu_observation_2023}.
Even in that case, the hopping between $B$ and $C$ orbitals is weak, as compared to their hopping to the $A$ orbital.
This motivates the examination in the limit with the direct hopping between $B$ and $C$ switched off, which becomes a fluxed dice lattice. In such a limit, one can find a singular flat band, which has exact flatness, but with a quadratic band touching point to the lower dispersive band~\cite{yang_fractional_2024}, and the higher dispersive band is always far detuned (by the energy of $\epsilon_A$), well negligible.
The quadratic band touching between the two lower bands can be minimally gapped by a weak nearest-neighbor repulsion, turning the singular band into a nearly flat Chern band. Surprisingly, two-band ED and all-orbital real-space DMRG calculations of the 3-orbital model unambiguously show the existence of FCI at $1/3$ and $2/3$ filling of the flat band~\cite{yang_fractional_2024}, even the gap is far insufficient to isolate it.

Here we go to the opposite and completely different limit, i.e. to obtain an isolated flat band by introducing anisotropy $\eta$ of single-particle hopping to fully gap the touching point [Fig.~\ref{Fig_band}(c)]. As already shown, with $\eta \neq 1$, one find a just complete set of non-orthogonal zero-modes [Fig. \ref{Fig_band}(d)], to span the exactly flat middle band, isolated from the lower dispersive band by a band gap of about $2(\eta-1)^2t^2/E_A$ [see Eq.~(\ref{Eq_gap})].
As expected, such an isolated band with exact flatness in this model must be topologically trivial \cite{francesco_cluster_2014}.
In fact, all three bands have zero Chern number in this anisotropic case.

\subsection{Reduction to a two-orbital honeycomb lattice with third nearest-neighbor hopping}

When $\epsilon_\mathrm{A}\gg t$, so that A orbital is virtually unoccupied, the lower two bands of the dice lattice could be very well reproduced by a minimal two-orbital model:
\begin{equation}
    \hat{H}_l=-R^\dagger R/\epsilon_\mathrm{A},
\end{equation}
where $R=(f(\mathbf{k}),g(\mathbf{k}))$ \cite{yang_fractional_2024}.
As shown in Fig.~\ref{Fig_band}(b), the retained B and C orbitals form a honeycomb lattice with longer-range hoppings introduced up to third nearest-neighbors, and an exemplary band structure is shown in Fig.~\ref{Fig_band}(c).
In this study, we set $t=1$ and $\epsilon_A=10$. In this parameter regime, the two-band model gives virtually identical description to the projected flat band (with Hartree-Fock renormalization) as the original three-orbital model. It further enables more efficient iDMRG simulations.

\section{Hartree-Fock band structure in the isolated band limit}
\label{sec_HF_band}
To investigate the many-body correlation at partial filling of this isolated flat band with $C=0$, an isotropic nearest-neighbor repulsion $\hat{H}_{\rm int}=\sum_{\langle l, m\rangle} U \hat{n}_l \hat{n}_m$ \cite{note1} is introduced between the B and C orbitals.
In the isolated flat band limit (c.f. definition below), the influence of the filled lower dispersive band can be well accounted by the Hartree-Fock mean field.
To perform the Hartree-Fock mean field, we express the $\hat{H}_0+\hat{H}_{\rm int}$ in the momentum space and decouple all four-fermion operator terms by evaluating all their possible contractions, and then get the solution in a self-consistent way (see more details in \ref{APP_Hartree}).

Figure.~\ref{Fig_HF} shows the self-consistent Hartree-Fock calculations at $\nu=1$ (flat band empty, lower dispersive band filled), based on $\hat{H}_0+\hat{H}_{\rm int}$.
With any given $U$, one observes a topological band inversion as function of $\eta$, from Chern band at the isotropic limit ($\eta = 1$) to isolated trivial flat band at smaller $\eta$. This is evident from the closure of band gap in the parameter space spanned by $\eta$ and interaction strength $U$ [Fig.~\ref{Fig_HF}(a)].

We define the isolated band limit as the regime where interaction with the filled band can only modestly affect the width of the flat band [Fig.~\ref{Fig_HF}(b)], but leaving its quantum geometry unchanged.
The change in quantum geometry can be quantified by the Bloch function fidelity $F(\bm k) \equiv | \langle \psi_{\rm HF} (\bm k) | \psi_{\rm flat} (\bm k) \rangle |$, where $\psi_{\rm HF}$ denotes the Hartree-Fock wavefunction of self-consistent Hartree-Fock calculations at $\nu=1$ and $\psi_{\rm flat}$ is the Bloch function in the non-interacting limit (see Eq.~(\ref{Eq_wave}) below). In Fig.~\ref{Fig_HF}(c), we plot $F_{\rm min} = \min_{\rm BZ} F(\bm k)$, the minimum fidelity value over the BZ, in the $\eta-U$ space. $F_{\rm min} \geq 0.99$ is used as a criteria to identify a parameter regime for the isolated band limit.
Note that this criterion provides an upper bound for $\eta$ at each fixed value of $U$, which is lower than the critical $\eta$ at which the gap between the two lower bands closes, as illustrated in Figs.~\ref{Fig_HF}(d)-\ref{Fig_HF}(f)  for several representative values of $U$.
In Sec. \ref{sec_ED}, we will focus on the region where the flat $C=0$ band is well isolated, and the projected single-band ED taking into account the Hartree-Fock renormalization can be well justified at its partial filling.

For most of this parameter regime at the isolated-band limit, the relation $\Delta_{\rm HF} \gg U > w$ is satisfied, with $w$ being the HF-renormalized band width of the isolated flat band, and $\Delta_{\rm HF}$ is the renormalized band gap between the  two bands.
In Figs.~\ref{Fig_BC}(a)-~\ref{Fig_BC}(b), the Berry curvature and the trace of quantum metric tensor from the Hartree-Fock renormalized flat band are shown, which shall be the same as the free system ($U=0$) at this isolated-band limit.
We find that, although the total Berry phase is 0, ${\rm Tr} \mathcal{G}$ and $\Omega_{\mathbf{k}}$ show a sharp peak at $\Gamma$, while being smooth and almost constant away from $\Gamma$ ($\Omega_{\mathbf{k}}$ in this region has the opposite sign to that at and around $\Gamma$).
Moreover, in Fig.~\ref{Fig_BC}(d), we show that the trace condition
${\rm Tr} \mathcal{G}-|\Omega_{\mathbf{k}}|=0$ for ``ideal bands'' in conventional Chern insulators almost holds in the majority of the BZ away from $\Gamma$~\cite{wang_exact_2021, Ledwith2022_ideal_flatband}.
In Fig.~\ref{Fig_BC}(e), the Hartree-Fock renormalized flat-band dispersion is shown.
Interestingly, we find that the energy dispersion $\epsilon_{\mathbf{k}}$ picks up a width that is comparable to $U$, exhibiting a minimum at the $\Gamma$ point where both the quantum metric and the Berry curvature have sharp peaks. 

Surprisingly, as will be shown later, at $\nu_\mathrm{F} = 2/3$ filling of such a single isolated $C=0$  band (overall filling $\nu = 5/3$ of the two-band model), we find unambiguous evidences for FCIs with quantized Hall conductivity of $\sigma_H= -\frac{1}{3}\frac{e^2}{h}$.
We will show clear evidence to attribute this to the coordination between the Berry curvature of this $C=0$ band and the quantum metric that affects the momentum space distribution of interacting carriers.
We note that there is a strong particle-hole dependence for this scenario at fractional fillings.
As a comparison, at $\nu_{\rm F}$ = 1/3 filling of this isolated flat band, we find no evidence of FCI (Fig. \ref{Fig_flatband-1-3}).

\begin{figure*}[htp!]
\centering
\includegraphics[width=1\textwidth]{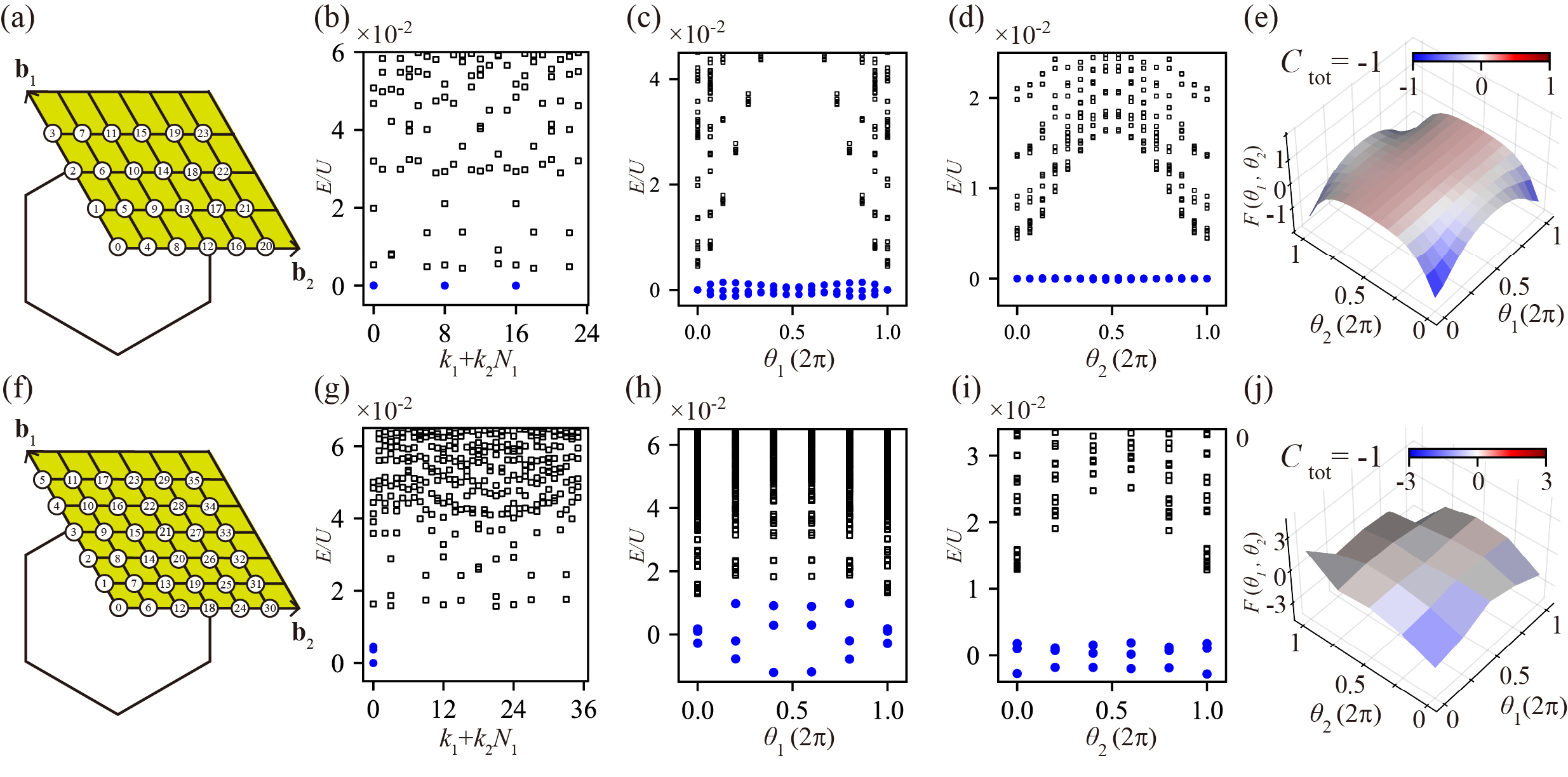}
\caption{
\textbf{ED results at $\nu_\mathrm{F}=2/3$ with $\eta=0.7$ and $U=0.001$.}
(a) Illustration of the $4\times6$ $k$-mesh, with all momenta indexed. (b) ED spectrum calculated at $\theta_1=0$ and $\theta_2=0$  for flat band filling $\nu_{\rm F}=2/3$, exhibiting three  quasi-degenerate ground states at K=0, 8 and 16, with a total Chern number of $-1$. (c) Spectral flow as a function of phase $\theta_1$, with $\theta_2=0$. (d) Spectral flow as a function of phase $\theta_2$, with $\theta_1=0$.
In the spectral flow figures here and hereafter, we have subtracted a background energy from the spectrum at every given phase $\theta$, for a better visualization of the spectral gap as a function of $\theta$ (c.f. \ref{APP_Background}).
(e) Distribution of many-body Berry curvature $F(\theta_1,\theta_2)$ (c.f. text).
(f)-(j) Similar plots for a $6\times6$ $k$-mesh.
}\label{Fig_ED6x6}
\end{figure*}

Before the numerical sections, we would like to point out another observed feature of the interesting quantum geometry in this isolated flat band. Its Bloch function can be expressed as:
\begin{equation}\label{Eq_wave}
| \psi_{\rm flat} \rangle=(0, g(\mathbf{k}),-f(\mathbf{k}))^T / \sqrt{|f(\mathbf{k})|^2+|g(\mathbf{k})|^2}.
\end{equation}
One can prove the trace of the quantum metric tensor ${\rm Tr} \mathcal{G}$ and the Berry curvature $\Omega_{\mathbf{k}}$ satisfy:
\begin{equation}\label{Eq_trace}
\text{Tr} \mathcal{G}+\Omega_{\mathbf{k}}=a^2,
\end{equation}
where $a$ denotes nearest neighbor bond length [c.f. Fig. \ref{Fig_band}(a)], as shown in Fig.~\ref{Fig_BC}(c).
If we flip the sign of the flux $\varphi$ in Fig. \ref{Fig_band}(a), $\Omega_{\mathbf{k}}$ changes sign but ${\rm Tr} \mathcal{G}$ is unchanged, and the above relation shall become $\text{Tr} \mathcal{G} -\Omega_{\mathbf{k}}=a^2$ instead.  

\section{Projected ED simulations of the isolated $C=0$ flat band}
\label{sec_ED}
\subsection{Robust FCI at $\nu_\mathrm{F}=2/3$ with $\sigma_\mathrm{H}=-\frac{1}{3}\frac{e^2}{h}$}

We begin with the $U=0.001$ and $\eta=0.7$ case.
The flat band characteristic with Hartree-Fock renormalization is shown in  Fig.~\ref{Fig_BC}(e).
As shown in Fig.~\ref{Fig_HF}(f), $\Delta_{\rm HF} \gg U$ is well satisfied, which justifies the single band ED calculations at fractional fillings.
This is further corroborated by comparing the many-body spectra of ED calculations with two bare bands and  ED with the single Hartree-Fock band, which show remarkable agreement for both the $\nu_\mathrm{F}=1/3$ and $\nu_\mathrm{F}=2/3$ fillings of the flat band (see \ref{APP_Comparison}).
In the following manuscript, the ED calculations are based on the single projected Hartree-Fock band if not specified.
This not only allows calculations with larger system size, but also sharpens the point that the FCI could emerge from an isolated $C=0$ band where band inversion is completely irrelevant.

By projecting the nearest-neighbor repulsive interaction onto the isolated flat band, we employ ED calculations on lattices with $N_s=N_1 \times N_2$ unit-cells put on a torus. In all ED calculations, a unique index $k=k_2 N_1 +k_1$ is assigned to a momentum $\mathbf{k}=\frac{k_1}{N_1} \mathbf{b}_1+\frac{k_2}{N_2} \mathbf{b}_2$ with $\mathbf{b}_1$ and $\mathbf{b}_2$ being the two reciprocal primitive vectors respectively [c.f. Figs. \ref{Fig_ED6x6}(a) and \ref{Fig_ED6x6}(f)].

The many-body spectra at $\nu_{\rm F} =  2/3$ filling of the flat band are shown in Figs. \ref{Fig_ED6x6}(b) and \ref{Fig_ED6x6}(g) for $4\times6$ [Fig. \ref{Fig_ED6x6}(a)] and $6\times6$ [Fig. \ref{Fig_ED6x6}(f)] system sizes, respectively.
Remarkably, we observe three nearly degenerate ground states separated by a finite spectrum gap from excited states.
The three-fold degeneracy of the ground state and their momenta are in agreement with those computed from the generalized Pauli principle for the 1/3 FCI on torus~\cite{Bernevig2008_jack, regnault_fractional_2011}. Taking the $6\times6$ system as an example, all three ground states are in the expected $\Gamma$ sector [Fig. \ref{Fig_ED6x6}(g)].

\begin{figure}[htp!]
\centering
\includegraphics[width=1\columnwidth]{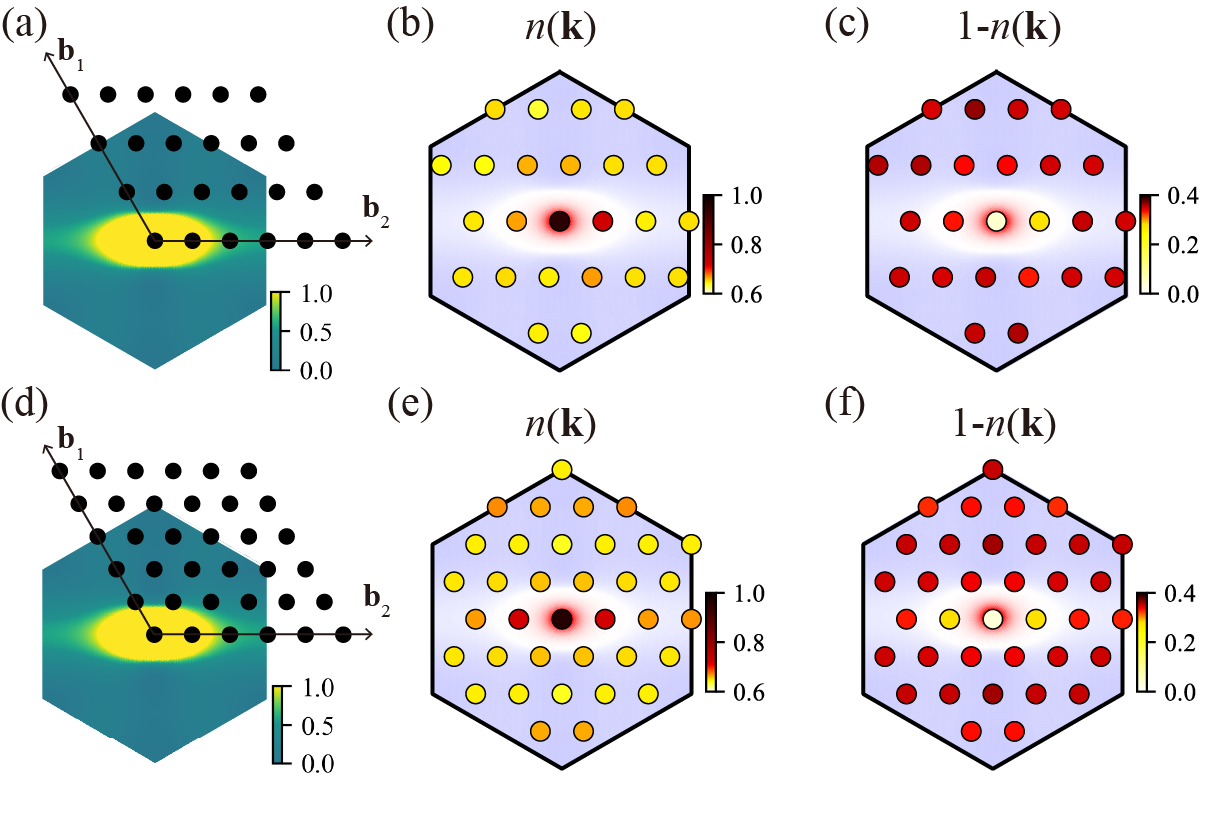}
\caption{\textbf{Momentum-space carrier distributions of the FCI.}
These results are obtained from ED simulations at $\nu_\mathrm{F}=2/3$ with $\eta=0.7$ and $U=0.001$.
 (a) Schematic of the $4\times6$ $k$-mesh, with BZ color coded with the value of ${\rm Tr} \mathcal{G}-|\Omega_{\mathbf{k}}|$. (b) Momentum-space electron distribution, $n(\mathbf{k})$, obtained as an average over the three ground states. (c) The corresponding hole density $1-n(\mathbf{k})$, which exhibits nearly zero weight at the $\Gamma$ point of the BZ. The background of the BZ in panels (b) and (c) is the Berry curvature of the flat band. (d)-(f) Similar plots but for the $6\times6$ $k$-mesh.
}\label{Fig_ED_comparison}
\end{figure}

\begin{figure*}[htp!]
\centering
\includegraphics[width=0.85\textwidth]{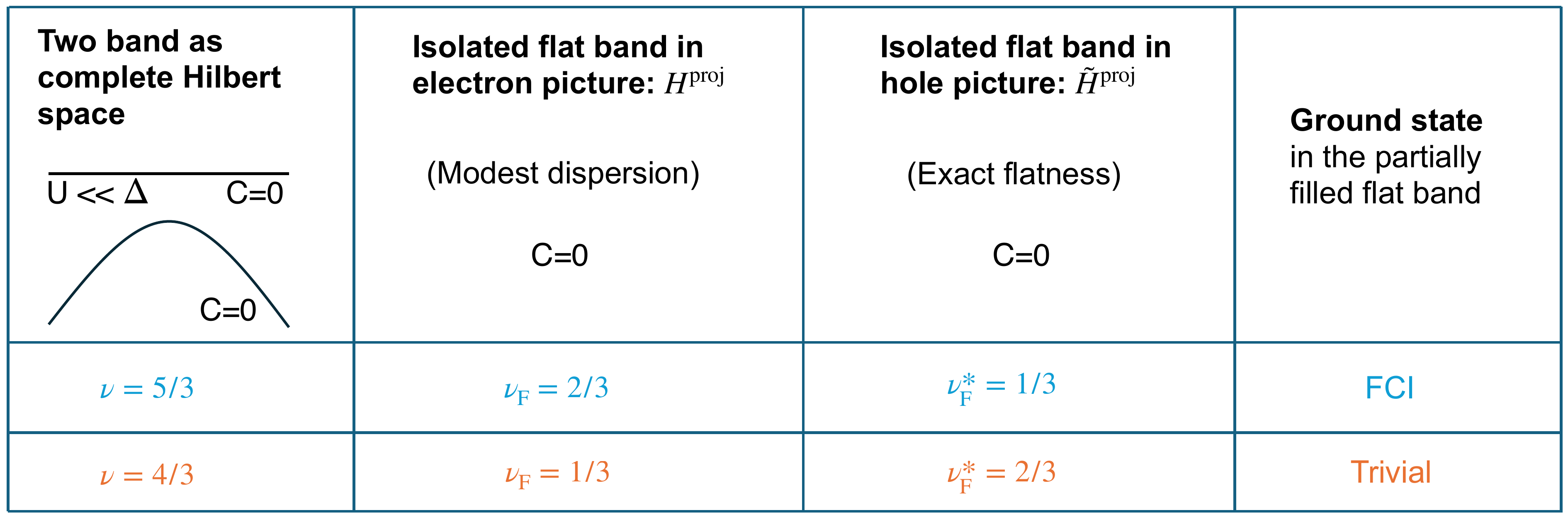}
\caption{
\textbf{Comprehensive conclusion of different pictures.}
}\label{Fig_table}
\end{figure*}

To probe the robustness of the spectrum gap, we insert fluxes by twisting the boundary conditions: $\hat{T}(N_i \mathbf{a}_i)|\Psi \rangle=e^{i \theta_i}|\Psi \rangle$, where  $\hat{T}(N_i \mathbf{a}_i)$ is the translation operator with $\mathbf{a}_i$ the primitive lattice vector and $|\Psi \rangle$ is the many-body state. The phase twist angles $\theta_1$ and $\theta_2$ play the role of magnetic fluxes, shifting the momentum from
$\frac{k_1}{N_1} \mathbf{b}_1+\frac{k_2}{N_2} \mathbf{b}_2$
to $(\frac{k_1}{N_1}+\frac{\theta_1}{2\pi N_1}) \mathbf{b}_1+(\frac{k_2}{N_2}+\frac{\theta_2}{2\pi N_2})\mathbf{b}_2$.
Notably, an evident energy gap consistently separates the ground states from the excited states in the spectral flow [Figs. \ref{Fig_ED6x6}(c), \ref{Fig_ED6x6}(d), \ref{Fig_ED6x6}(h) and \ref{Fig_ED6x6}(i)].
The robust gap, and the 3-fold degenerate ground states with the characteristic momenta, strongly suggest the existence of FCI.

To further identify the topological nature,  we calculate the many-body Chern number defined as an integral of many-body berry curvature $F(\theta_1, \theta_2)$ over the twisted boundary phase space
\begin{equation}
C=\frac{1}{2 \pi} \int_0^{2 \pi} d \theta_1 \int_0^{2 \pi} d \theta_2 F\left(\theta_1, \theta_2\right),
\end{equation}
where $F(\theta_1, \theta_2)=\text{Im}[\langle \partial_{\theta_1}\Psi|\partial_{\theta_2}\Psi\rangle-\langle\partial_{\theta_2} \Psi|\partial_{\theta_1} \Psi\rangle]$. The many-body Chern number links to the Hall conductance via $\sigma_{\rm H}=\frac{e^2}{h} C_{\rm ave}$ with $C_{\rm ave}=C_{\rm tot}/N_{\rm g}$, where $C_{\rm tot}$ is the sum of Chern numbers over the $N_{\rm g}$ degenerate ground states.
In practice, we take a $N_{\theta} \times N_{\theta}$ grid.
The many-body Berry curvature $F(\vec{\theta})=F(\theta_1, \theta_2)$ for each square plaquette
is then obtained by calculating the consecutive wave function overlaps $\langle\Psi(\vec{\theta})|\Psi(\vec{\theta'})\rangle$ around this square plaquette.
Specifically,
\begin{equation}
F(\vec{\theta})=\arg \left[U_1(\vec{\theta}) U_2(\vec{\theta}+\delta \theta_1)U_1^{-1}(\vec{\theta}+\delta \theta_2) U_2^{-1}(\vec{\theta})\right]
\end{equation}
with
\begin{equation}
U_i(\vec{\theta}) = \frac{\langle \Psi(\vec{\theta}) | \Psi(\vec{\theta} + \delta\theta_i) \rangle }
{\left| \langle \Psi(\vec{\theta}) | \Psi(\vec{\theta} + \delta\theta_i) \rangle \right|}.
\end{equation}
Here $\delta\theta_1=(2\pi/N_{\theta},0)$ and $\delta\theta_2=(0,2\pi/N_{\theta})$

The distribution of the many-body Berry curvature in a $15 \times 15 $ grid for the $4\times6$ system and in a smaller grid for the $6\times6$ system, summed over the three ground states, are shown in Figs. \ref{Fig_ED6x6}(e) and \ref{Fig_ED6x6}(j), respectively.
Remarkably, the total Chern number is exactly $-1$ and thus the Hall conductivity is $\sigma_{\rm H}=-\frac{1}{3}\frac{e^2}{h}$.
In this sense, we have identified the emergence of an FCI with quantized Hall conductivity of $\sigma_{\rm H} = -\frac{1}{3}\frac{e^2}{h}$ at $\nu_\mathrm{F} = 2/3$ filling of the isolated $C=0$ flat band.

Furthermore, at this parameter ($\eta=0.7$ and $U=0.001$), we have considered ED simulations of other clusters with different sizes and meshes, where the robustness of this FCI is consistently verified in all calculations [c.f. \ref{App_Supporting}].

To understand the relation between filling and Hall conductivity, which is different from the FCI from any Chern band, we further computed the momentum-space electron distributions and corresponding hole distributions for both the $4\times6$ and $6\times6$ $k$-meshes [Fig.~\ref{Fig_ED_comparison}]. The electron distributions, $n(\mathbf{k})$ averaged over the three nearly degenerate ground states, are shown in Figs.~\ref{Fig_ED_comparison}(b) and ~\ref{Fig_ED_comparison}(e).
Consistent with the Hartree-Fock band dispersion, the electron density peaks around $\Gamma$ point where the Hartree-Fock band shows a minimum.
Correspondingly, the holes predominantly occupy regions away from the $\Gamma$ point, as evidenced by the plot of $1-n(\mathbf{k})$ in Figs.~\ref{Fig_ED_comparison}(c) and ~\ref{Fig_ED_comparison}(f), where the Berry curvature is positive and nearly uniform, and ${\rm Tr} \mathcal{G}-|\Omega_{\mathbf{k}}| \approx 0$.
In this context, it is the hole carrier that plays the dominant role of the FCI by occupying the region of the BZ which is analogous to that of an ideal Chern band, explaining the observed Hall conductivity.

\subsection{Coordinating role of the quantum metric and the Berry curvature}

In the previous section, we demonstrated that the holes in the FCI are distributed away from the $\Gamma$ point, where the Berry curvature is positive and nearly uniform.
At first glance, one may simply attribute this carrier distribution to the energy minimum of the Hartree-Fock band at the $\Gamma$ point [Fig.~\ref{Fig_BC}(e)].
However, in this section, we argue that it is the intrinsic quantum geometry that underlies this unexpected FCI, with a coordination between the quantum metric and the Berry curvature.

For the ED calculations presented above, we consider the electron filling $\nu_\mathrm{F}=2/3$ of the flat band (corresponding to a total filling $\nu=5/3$), and the calculations are based on the projected Hamiltonian (denoted as $H^{\rm proj}$) that accounts for the effective band dispersion from the Hartree-Fock renormalization [Fig.~\ref{Fig_BC}(e)].
As shown in \ref{APP_HF_renormalized}, this dispersion from Hartree-Fock renormalization can be expressed in terms of the quantum metric of the filled dispersive band, which is identical to that of the flat band.

Switching to the hole picture, we note that the particle-hole transformation of the projected Hamiltonian $H^{\rm proj}$ 
results in an emergent kinetic energy~\cite{Abouelkomsan2023_metric_induce} that exactly cancels the effective dispersion resulting from the Hartree-Fock correction from the lower dispersive band. A more detailed discussion of the emergent kinetic energy and the Hartree-Fock-corrected  band dispersion is provided in \ref{APP_HF_renormalized}.
As a result, the particle-hole transformation yields the projected Hamiltonian $ \tilde{H}^{\rm proj}$ in the hole picture, whose effective band dispersion is exactly flat.
We have checked that the ED calculations of $ \tilde{H}^{\rm proj}$ at the hole filling  $\nu_\mathrm{F}^\ast=1/3$ on the $4\times6$ and $6\times6$ clusters yield the same many-body spectra as those shown in Fig. \ref{Fig_ED6x6}(b) and Fig. \ref{Fig_ED6x6}(g), respectively, and the corresponding hole distributions match those shown in Fig. \ref{Fig_ED_comparison}(c) and Fig. \ref{Fig_ED_comparison}(f).
The hole distribution observed in this calculation aligns with previous studies that, in flat bands with repulsive interactions, the carriers preferentially occupy regions of the BZ where quantum metric is minimized \cite{Lauchli2013_FCI_dispersion, Ji2024_metric_induced_holedispersion}.

The table in Fig.~\ref{Fig_table} concludes the scenario here from different perspectives at different fillings,
with equivalent underlying physics in the electron and hole representations, albeit the apparent particle-hole asymmetry in the projected band.
As suggested by the momentum-space carrier distribution, the coordination of the quantum metric and the Berry curvature allows the interacting carriers to preferentially occupy a BZ region with nearly ideal quantum geometry in an otherwise topologically trivial flat band, enabling the formation of FCI.
This could be a possible scenario for exploring FCIs in $C=0$ flat bands in general.

\begin{figure}
\centering
\includegraphics[width=1\columnwidth]{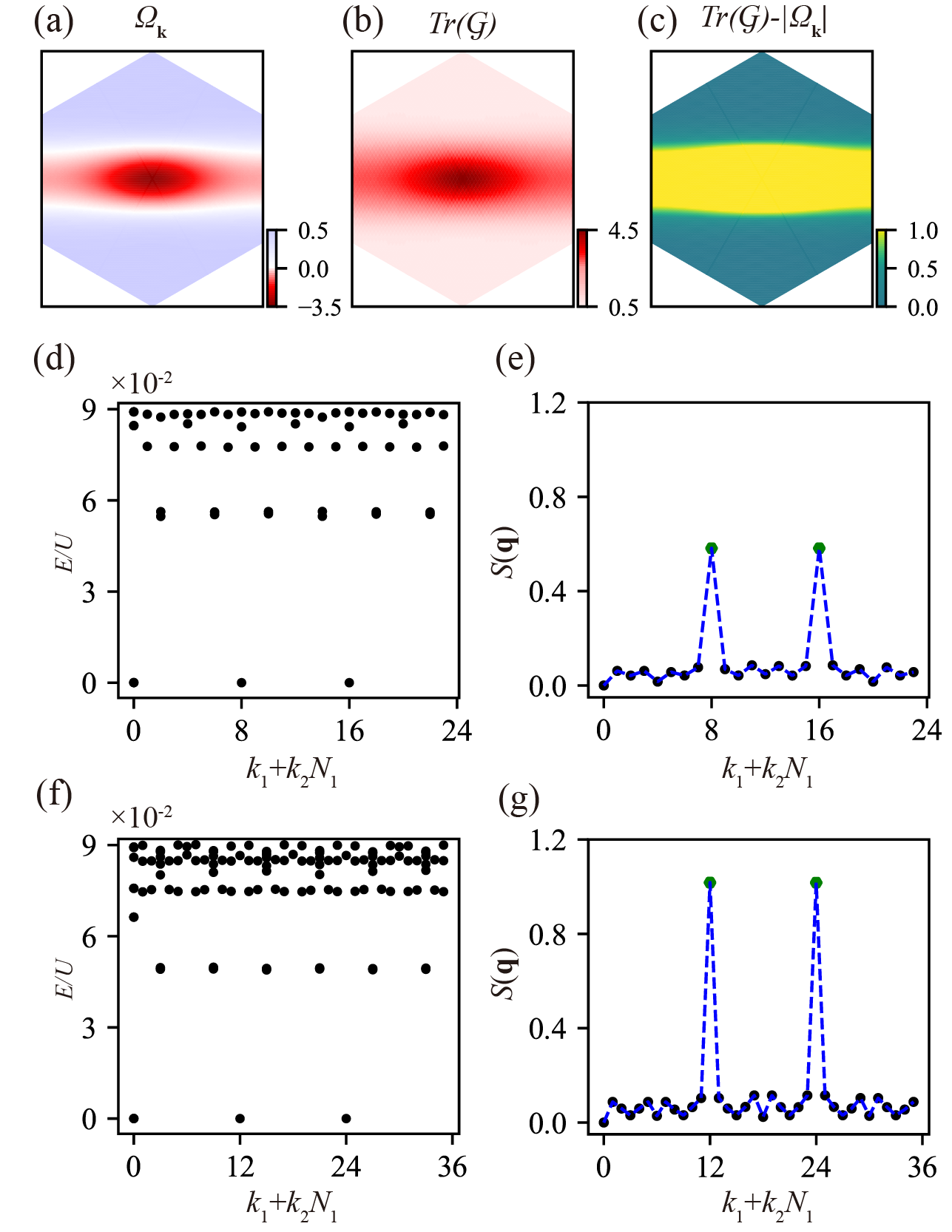}
\caption{ \textbf{Structure factors for the $4\times6$ and $6\times6$ $k$-meshes at $\nu_{\rm F}=2/3$ with $\eta=0.2$ and $U=0.001$.} (a) Berry curvature $\Omega_{\mathbf{k}}$, (b) trace of quantum metric tensor $\mathcal{G}$, and (c) ${\rm Tr} \mathcal{G}-|\Omega_{\mathbf{k}}|$ of the flat band.
(d) ED spectrum from the $4\times6$ $k$-mesh. (e) Structure factor distribution of the lowest state at K=0 sector from the $4\times6$ $k$-mesh. (f) ED spectrum of the $6\times6$ $k$-mesh. (g) Structure factor distribution of the lowest state at K=0 sector from the $6\times6$ $k$-mesh.
}\label{Fig_cdw}
\end{figure}

In this scenario, we have observed two features.
1. There is a strong particle-hole asymmetry, and for the type of carriers that see an exact flat band, a lower density is more favorable for the FCI. For the example here, the projected band for holes is exactly flat. If we raise the hole filling to $\nu_\mathrm{F}^\ast=2/3$, we did not find evidence of FCI [Fig. \ref{Fig_flatband-1-3}(b)].
2. The sharpness of the peak of quantum metric and Berry curvature also matters. As will be discussed in the next subsection, when the peaks at $\Gamma$ become less sharp as we increase the anisotropy (c.f. Fig.~\ref{Fig_cdw}), the ground state transitions from FCI to the inter-chain CDW.

\subsection{Inter-chain CDW in the small-$\eta$ limit}
As discussed above, the emergence of FCI from this $C=0$ band originates from the fact that the hole carriers tend to avoid the peaks of quantum metric and Berry curvature at $\Gamma$, while occupying the rest of the BZ with more ``ideal'' conditions.
An interesting and natural question is to examine the effect of sharpness of the quantum metric and Berry curvature peaks.
At the single-particle level, we observe that when decreasing $\eta$, the sharpness of these quantum geometric quantities at $\Gamma$ also decreases. Therefore, we choose $\eta=0.2$ and the same interaction strength $U=0.001$ to explore the ground state.

As shown in Figs. \ref{Fig_cdw}(a)-\ref{Fig_cdw}(c), the quantum metric and Berry curvature peaks at $\Gamma$ are less sharp as compared to the case of $\eta=0.7$ studied above, and the region of the BZ with ${\rm Tr} \mathcal{G}-|\Omega_{\mathbf{k}}| \approx 0$ gets smaller.
The spectra of the ED simulations under the $4\times6$ and $6\times6$ k-meshes are shown in Figs. \ref{Fig_cdw}(d) and \ref{Fig_cdw}(f), respectively. Although we still observed 3 low-lying states, for the $6\times6$ mesh, the momenta of ground states (only one state at $\Gamma$) does not meet the generalized Pauli principle for the 1/3 FCI~\cite{Bernevig2008_jack, regnault_fractional_2011}, suggesting that the ground state is no longer the FCI found at $\eta=0.7$. In this sense, one possible candidate state is the CDW.

Therefore, to examine the possible charge order, we calculate the structure factors defined as
\begin{equation}
S(\mathbf{q})=\frac{1}{N_s}\sum_{\alpha\beta}(\left\langle\hat{\rho}_\alpha(\mathbf{q}) \hat{\rho}_\beta(-\mathbf{q})\right\rangle-\left\langle\hat{\rho}_\alpha(\mathbf{0}) \rangle\langle \hat{\rho}_\beta(\mathbf{0})\right\rangle \delta_{\mathbf{q},0}),
\end{equation}
where $\hat{\rho}_\alpha(\mathbf{q})=\hat{\alpha}^{\dagger}_{\mathbf{k}}\hat{\alpha}_{\mathbf{k}+\mathbf{q}}$ with $\alpha,\beta=A,B,C$.
We calculate $S(\mathbf{q})$ from one of the ground states (at $\Gamma$), and the results from the $4\times6$ and $6\times6$ $k$-meshes are shown in Figs. \ref{Fig_cdw}(e) and \ref{Fig_cdw}(g), respectively.
Sharp Bragg peaks that increase with system sizes are observed, confirming that the ground state here is a CDW.
More specifically, from the specific wave vector, the translation symmetry is broken in one direction of the primitive vectors and the unit cell is tripled.
This specific CDW order is easier to understand if we consider the $\eta=0$ limit, where the lattice becomes disconnected into arrays of one-dimensional chains, and this phase is just an inter-chain CDW and gradually gives way to the FCI at larger $\eta$ with sharper quantum geometry.
This CDW order is further verified in the large-scale iDMRG calculations, as discussed in section \ref{sec_dmrg_A}.

\begin{figure}
\centering
\includegraphics[width=1\columnwidth]{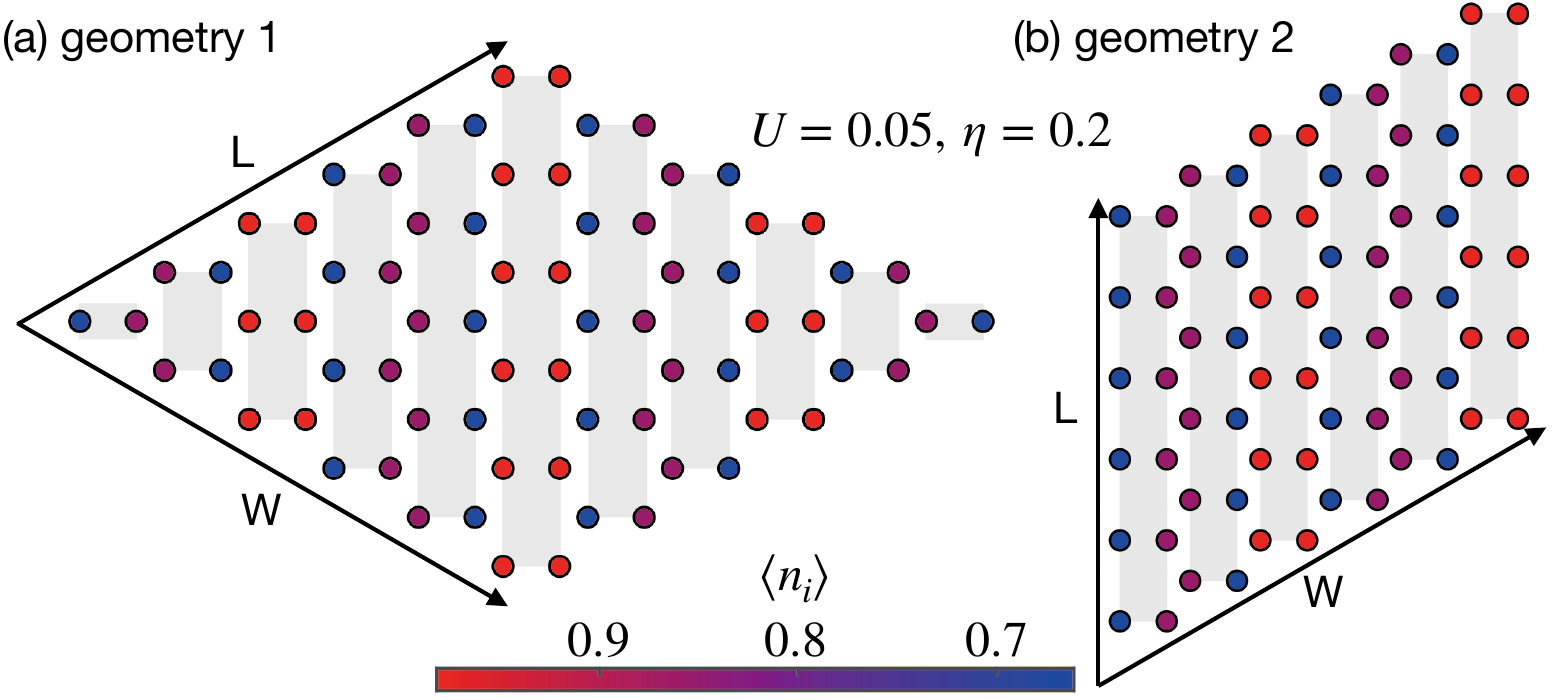}
\caption{\textbf{Real-space geometries of iDMRG simulations and corresponding CDW patterns.} Here, two different geometries are considered in the iDMRG simulations, as shown in panels (a) and (b) respectively, with W being the circumference of the cylinder and L being in the thermodynamic limit.
The grey shadow refers to the disconnected chains of the model at $\eta\rightarrow0$.
The color at each lattice site refers to the electron density from iDMRG results at $\nu=5/3$ with $U=0.05,\ \eta=0.2$.
In geometry 1, neither real-space primitive vector is along the disconnected chains but both (W, L) go across the chains, but the measured density pattern is still consistent with the inter-chain CDW and the period of both directions is tripled.
In geometry 2, the length (L) of the infinite cylinder is along the chain and the density is invariant, and the inter-chain CDW triples the period of density along the circumference (W).
}\label{Fig_dmrg_geometry}
\end{figure}

\section{iDMRG results of the two-band model}
\label{sec_DMRG}
In this section, we focus on the iDMRG simulations~\cite{White1992_dmrg, mcculloch2008_idmrg} of the two-orbital honeycomb lattice model, as shown in Fig.~\ref{Fig_band}(b).
Throughout this section, we consider fixed interaction $U=0.05$ at different fillings and tune $\eta$, as smaller $U$ might take much larger bond dimension of iDMRG simulations for good convergence.
This larger $U$ as compared to that in the ED simulations has limited the isolated band limit of the $C=0$ flat band to significantly smaller $\eta$ range [c.f. Fig.~\ref{Fig_HF}(d)].


\begin{figure*}[htp!]
\centering
\includegraphics[width=\textwidth]{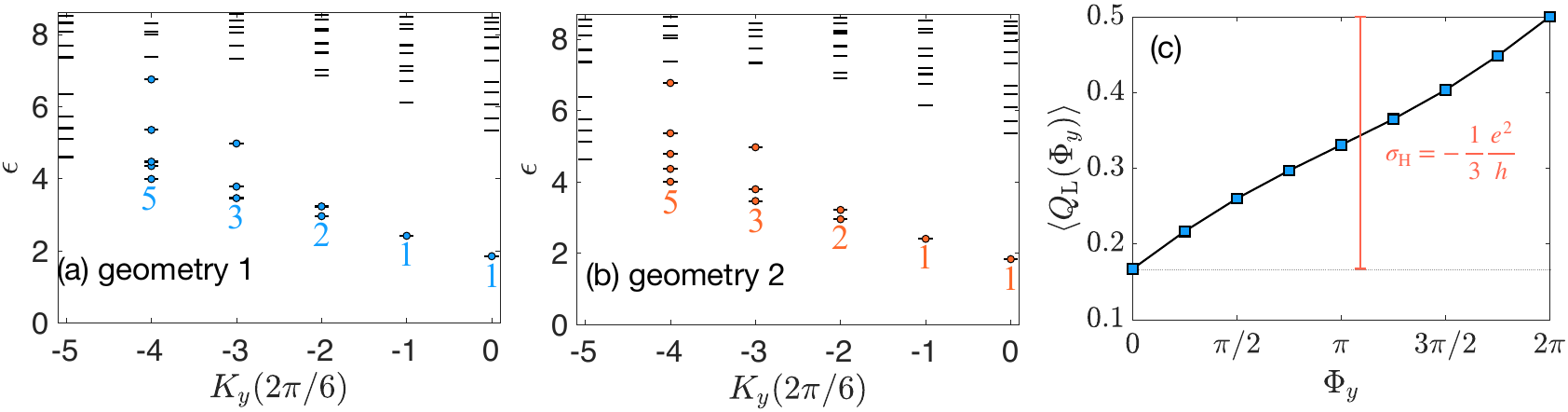}
\caption{\textbf{Evidence of FCI at $U=0.05,\ \eta=0.3,\ \nu=5/3$.} The momentum-resolved entanglement spectrum ($\epsilon$) in the charge sector with the lowest $\epsilon$ from geometry 1 (a) and 2 (b).
The results from both geometries show the characteristic counting $\{
1,1,2,3,5,... \}$ of the edge conformal field theory, confirming the topological nature of this Laughlin state.
(c) The charge pumping from geometry 1 as an example. After adiabatically inserting $2\pi$ flux, $1/3$ charge is pumped from right to left, supporting the fractionally quantized $\sigma_\mathrm{H}=-\frac{1}{3}\frac{e^2}{h}$.
}\label{Fig_fci_evidence_dmrg}
\end{figure*}

\subsection{Real-space geometries and the CDW pattern}\label{sec_dmrg_A}
At $\eta\rightarrow 0$, the non-interacting lattice would be disconnected into 1-dimensional chains.
Therefore, we consider two real-space geometries as shown in Fig.~\ref{Fig_dmrg_geometry}.
In geometry 1, either primitive vectors would not go along the disconnected chains, but across them, which is similar to the $k$-mesh 1 in ED simulations (refer to \ref{APP_Competition} for the definition of the $k$-meshes).
In geometry 2, we take one vector along the chains while the other one across them, which is similar to the $k$-meshes 2 and 3 in ED simulations.
In both cases, we take the finite circumference of the infinite cylinder to be across the chains, such that the direction with infinite length of the cylinder is across or along the chains for geometry 1 and 2, respectively.
We have considered up to $N_\mathrm{W}=6$ unit cells (12 lattice sites) for the finite circumference (W), and different numbers of sites for the iDMRG unit cell in the infinite L direction are considered for convergence.
Therefore, we do not suffer from finite-size effect in the iDMRG simulations regarding the compatibility of system sizes and the inter-chain CDW.
The $U(1)$ charge conservation symmetry is implemented in the simulations and the bond dimension is up to $D=3000$ for the well-converged results with the maximum truncation error on the order of $10^{-6}$.

To have a clearer demonstration of the inter-chain CDW order, the real-space electron densities from both geometries at $\nu=5/3$ (flat band filling $\nu_F=2/3$) are shown in Fig. \ref{Fig_dmrg_geometry} as well, for $U=0.05,\ \eta=0.2$. In order to show explicitly translation-symmetry breaking, we do not implement the translation symmetry of the Hamiltonian along $W$ in the simulations here.
In geometry 2 [Fig. \ref{Fig_dmrg_geometry}(b)], the real-space configuration is consistent with the structure factor of ED results that the electron density is invariant along the chains while the translation period is tripled across the chains (along the circumference W).
In geometry 1, although neither L or W is along the disconnected chains, the converged real-space configuration is almost the same as that from geometry 2, still in agreement with the inter-chain CDW order.
This further suggests, at small $\eta$ (large anisotropy), the CDW phase should be the ground state of $\nu=5/3$ in the thermodynamic limit.

\subsection{Evidence of FCI from the $C=0$ band in the isolated band limit}

Having introduced the geometries of the iDMRG simulations and the details of the small-$\eta$ CDW phase, we now turn to the robust evidence of the FCI phase at $\nu=5/3$.
As we consider fixed $U=0.05$, we will focus on the $\eta=0.3$ for demonstration of the FCI, where the gap between the two bands is much larger than $U$,  ensuring the $C=0$ flat band is in the isolated band limit [Fig.~\ref{Fig_HF}(d)], namely the Bloch function fidelity is lower bounded by $F_\mathrm{min}>0.99$.
Nevertheless, we do not implement any band projection in iDMRG as is done for the ED, but directly simulate the interacting two-orbital Hamiltonian with these parameters.
The iDMRG simulations iteratively optimize the matrix-product-state representation of the ground state through Schmidt decomposition of the system into two half infinite cylinders and truncates it at a given bond: $|\Psi\rangle=\sum_\alpha \Lambda_\alpha|\alpha_\mathrm{L}\rangle |\alpha_\mathrm{R}\rangle$.
This decomposition provides a convenient access to the entanglement properties and the entanglement spectrum (ES) could be obtained from $\Lambda_\alpha^2=e^{-\epsilon_\alpha}$~\cite{Cincio2013_topological_order, Zaletel2013_topological_characterization}.
The momentum-resolved ES in the charge sector with the lowest $\epsilon_\alpha$ (we have implemented $U(1)$ charge conservation symmetry of the Hamiltonian that also holds for the reduced density matrix from the Schmidt states) from geometry 1 and 2 are shown in Figs. \ref{Fig_fci_evidence_dmrg}(a) and \ref{Fig_fci_evidence_dmrg}(b), respectively.
From both geometries, the ES show the characteristic edge-mode
counting $\{ 1,1,2,3,5,... \}$, in agreement with the edge conformal field theory, which further provides the smoking gun evidence of the topological order~\cite{Li-Haldane2008_ES, Regnault2015_ES}.

The charge pumping result is shown in Fig. \ref{Fig_fci_evidence_dmrg}(c) for geometry 1 as an example.
After adiabatically inserting $2\pi$ flux, a fractionally quantized $1/3$ charge is pumped from the right to the left of the system, which is exactly in agreement with the ED calculations that this FCI state has a fractionally quantized $\sigma_\mathrm{H}=-\frac{1}{3}\frac{e^2}{h}$ at $\nu=5/3$~\cite{Grushin2015_FCI}.
These results further support the existence and robustness of the FCI from the isolated trivial flat band.

\subsection{Quantum phase diagram at $\nu=1$ and $\nu=5/3$}

\begin{figure}
\centering
\includegraphics[width=1\columnwidth]{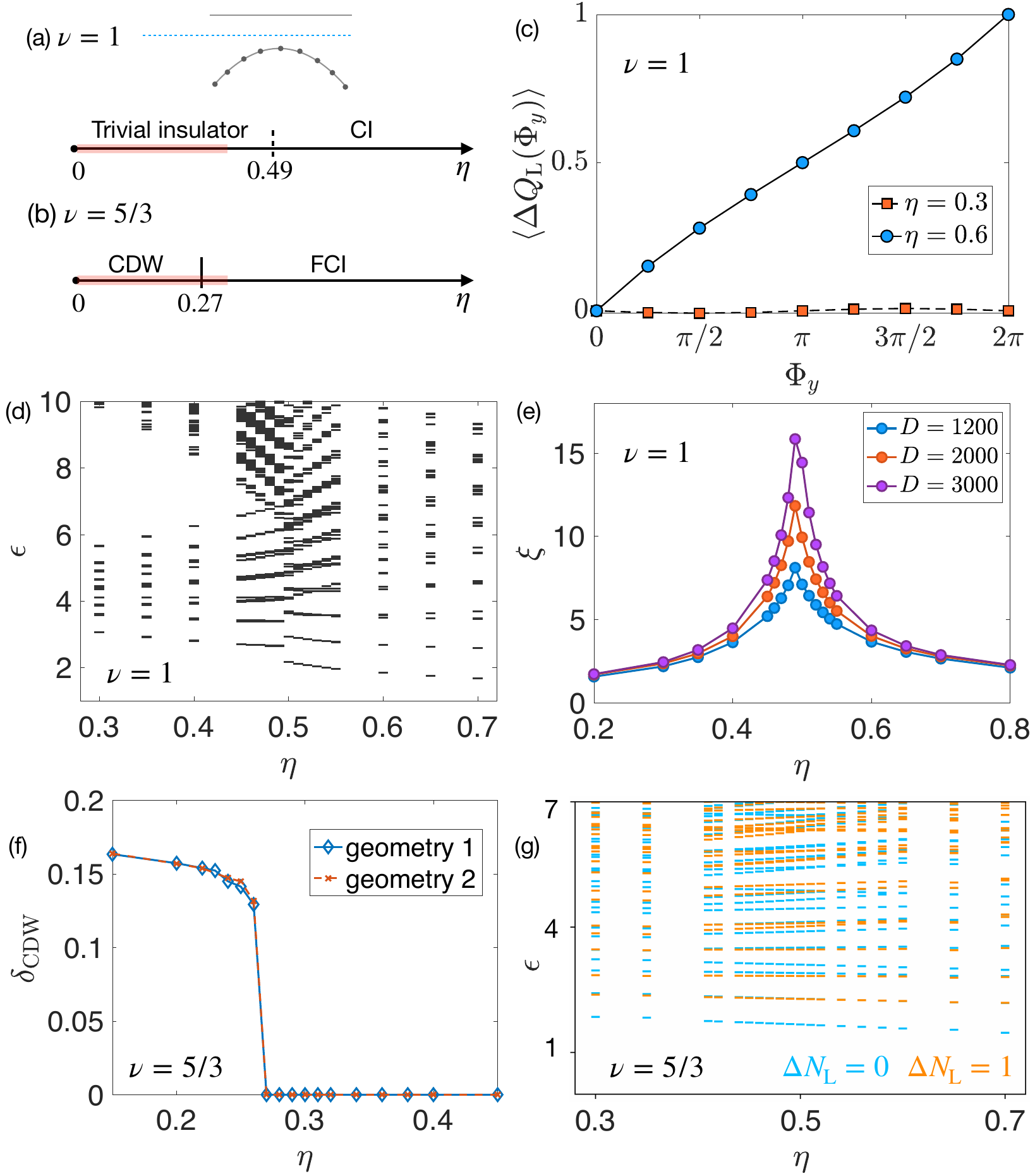}
\caption{\textbf{Phase diagram and transitions at $U=0.05$.} Quantum phase diagrams with $U=0.05$ are shown for (a) $\nu=1$ and (b) $\nu=5/3$, with the phase boundaries supported by both geometries. The shaded region refers to the isolated band limit.
At $\nu=1$ (the schematic of this filling is shown above the phase diagram with the lower dispersive band fully filled), the topological transition happens at $\eta\sim0.49$ between the trivial band insulator and the CI phases. The full ES ($\epsilon$), and the correlation length ($\xi$) from different bond dimensions of the iDMRG simulations across the transition are shown in panels (d) and (e) respectively, supporting the gap closing at the transition point.
The charge pumping of $\eta=0.3,\ 0.6$ at $\nu=1$ are shown in panel (c), with one electron pumped for the CI after adiabatically inserting $2\pi$ flux, while no charge pumped for the trivial insulator. Panels (c)-(e) are from geometry 1 for example.
At $\nu=5/3$, the CDW-FCI transition happens at $\eta\sim0.27$, as shown from the sharp change of the CDW order parameter around the transition point in panel (f).
The ES (of the same topological sector) of two lowest charge sectors at $\nu=5/3$ from geometry 1 is shown in panel (g). $\Delta N_\mathrm{L}=N_\mathrm{L}-N_\mathrm{L}^0$, with $N_\mathrm{L}^0$ being the charge sector with the lowest level of $\epsilon$.  Denser data points are taken around the transition point of the integer filling ($\eta\sim0.49$).
In the same topological sector, the ES of the FCI from trivial band smoothly evolves into that of the FCI from Chern band (larger $\eta$), suggesting the FCIs from different origins are adiabatically connected.
}\label{Fig_dmrg_phase_diagram}
\end{figure}

To further understand this exotic FCI from the $C=0$ flat band, we still consider the fixed interaction $U=0.05$ while tune $\eta$ at different fillings.
We perform the iDMRG simulations at the integer filling $\nu=1$ as well, with the same geometries as are performed at the fractional fillings.
The quantum phase diagram at this integer filling is shown in Fig.~\ref{Fig_dmrg_phase_diagram}(a). The results from both geometries indicate a transition point at $\eta\sim0.49$ at the interaction strength $U=0.05$, also consistent with the Hartree-Fock calculations at this filling. The ground state at $\nu=1$ is a trivial band insulator at $\eta<0.49$ but an integer Chern insulator (CI) at $\eta>0.49$.
We also show the charge pumping results for these two phases in Fig.~\ref{Fig_dmrg_phase_diagram}(c). After adiabatically inserting $2\pi$ flux, there is exactly one electron pumped in the CI phase, while no response in the trivial insulator.
The full ES [Fig. \ref{Fig_dmrg_phase_diagram}(d)] shows a gap closing around $\eta\sim0.49$ and the correlation length [Fig. \ref{Fig_dmrg_phase_diagram}(e)] tends to diverge with increasing bond dimensions, suggesting the (at least almost) gapless transition point at $\eta\sim0.49$.

We then turn to the fractional filling $\nu=5/3$ and try to determine the phase boundary between the CDW and the FCI phases at $U=0.05$.
As the real-space density configurations are almost the same inside the CDW phase from both geometries, as shown in Fig.~\ref{Fig_dmrg_geometry}, and the densities are uniform in the FCI phase, we simply define the CDW order parameter here as $\delta_\mathrm{CDW}=\langle n\rangle_\mathrm{max}-\bar{n}$ to detect the phase boundary.
Since the average density is $\bar{n}=5/6$, this defined order parameter would range from $0$ to $1/6$.
As shown in Fig. \ref{Fig_dmrg_phase_diagram}(f), results from both geometries consistenly give the phase boundary around $\eta\sim0.27$, where the CDW order jumps to $0$ as  $\eta$ increases from small to large values.

As discussed earlier, at $U=0.05$ and $\eta>0.49$, the ground state of $\nu=1$ is a CI with $\sigma_\mathrm{H}=-\frac{e^2}{h}$, and the ground state of $\nu=5/3$ is an FCI (we have checked the Hall conductivity is still  $\sigma_\mathrm{H}=-\frac{1}{3}\frac{e^2}{h}$).
In this large $\eta$ regime well beyond the isolated band limit, the interaction renormalization gives rise to a nearly flat Chern band in the first place, which sustains the FCI at the partial filling. This is completely different from the the FCI hosted by the isolated trivial band at smaller $\eta$.
In the Chern band regime, apart from the scenario of a $\nu_F^\ast=1/3$ hole FCI,
the FCI here could also be understood from a fully occupied $C=-1$ band plus a $2/3$-filled $C=1$ band that hosts a 2/3 electron FCI.
Interestingly, as shown in Fig.~\ref{Fig_dmrg_phase_diagram}(g) with denser data points around $\eta\sim0.49$, the ES (of the same topological sector) of the FCI from isolated trivial band smoothly and continuously evolves into that of the FCI from the Chern-band scenario.
This supports that the FCIs from different origins (isolated trivial band or Chern band) are adiabatically connected, which further suggests that the understanding of FCIs should not be limited to but beyond the single-particle band topology.


\section{Conclusion and discussion}
\label{sec_discussion}

Our finding provides an example that the FCI could emerge from isolated trivial band. More importantly, we demonstrate a convincing scenario that, in an exactly flat $C=0$ band, due to the coordination of quantum metric and Berry curvature, the partially filled carriers might prefer to avoid the sharp peak of quantum metric and Berry curvature while occupying the rest of the BZ.
Given the nearly ``ideal'' conditions away from the peaks of the quantum metric and Berry curvature ($\Gamma$ point in our case), the FCI emerges.
This scenario has a prominent particle-hole dependence, which is supported by the FCIs at $\nu_\mathrm{F}=2/3$ and the trivial states at $\nu_\mathrm{F}=1/3$.
According to our analysis, the sharpness of the quantum geometry might also be important, since the ground state at $\nu_\mathrm{F}=2/3$ becomes the trivial CDW at smaller $\eta$ with less sharp peaks in the quantum geometric quantities.

Besides, the combination of the obtained fractionally quantized Hall conductivity, ground state degeneracy, and the filling factor also suggests that the emergence of the FCIs from the isolated trivial flat band is not relevant to the band-folding Hall-crystal scenarios~\cite{ Dong2024_AHC_graphene, Lu2025_EQAH_graphene, Waters2025_CI_fractional_filling_pentalayer_graphene, Sheng2024_QAHC_mote2, Lu2025_FQAHC}.
In such fractional Hall crystals, the CDW order leads to the formation of renormalized minibands in the folded BZ which can have nontrivial band topology, and partial filling of such Chern minibands leads to the formation of FQAH states~\cite{Lu2025_FQAHC}. This can not be the case here, for two-fold reasons. First, from the band-folding perspective, since the CDW here (at small $\eta$) triples the unit cell and the original filling of the isolated band is $2/3$, the effective filling in the folded BZ should be an integer and the corresponding total Chern number of the filled minibands should also be an integer, not possible for the observed fractionally quantized Hall conductivity~\cite{Xie2021_topological_CDW_TBG, Polshyn2022_topological_CDW, Su2025_TEC_graphene, Sheng2024_QAHC_mote2, Lu2025_FQAHC}. Second, if there is coexisting CDW, the ground-state degeneracy would be enlarged~\cite{Kourtis2014_pinball, kourtis_symmetry_2018, Lu2024_FQAHS, Lu2025_gapless_roton}, being the product of the topological and CDW degeneracies, but we only observe 3-fold degeneracy in the FCIs here. Therefore, we conclude that the emergence of FCIs here is a radically new scenario, where band topology is not a precondition for the emergence of many-body topological order.

Acknowledgments: The authors thank Jie Wang for helpful discussions. The work is supported by the National Natural Science Foundation of China (No. 12425406), Research Grant Council of Hong Kong (AoE/P-701/20, HKU SRFS21227S05), and New Cornerstone Science Foundation. The authors thank Beijing PARATERA Tech Co., Ltd. (https://cloud.paratera.com) for providing HPC resources that supported the research results reported in this paper.

Notes: W.Y. conceived the model, designed and supervised the research. Z.L. carried out  ED calculations with inputs from H.L., W.Q.Y, and D.Z.
H.L. performed the DMRG calculations.
W.Q.Y. performed the Hartree-Fock calculations. All authors discussed and analyzed the results. Z.L., H.L., and W.Y. wrote the manuscript with inputs from all authors.

\appendix
\def\thesection{Appendix \Alph{section}} 

\def\@seccntformat#1{%
  \ifcsname prefix@#1\endcsname
    \csname prefix@#1\endcsname
  \else
    \csname the#1\endcsname\quad
  \fi}
\def\prefix@section{Appendix \Alph{section}: } 
\renewcommand{\theequation}{\Alph{section}\arabic{equation}}

\section{Hartree Fock mean-field}
\label{APP_Hartree}
To perform the Hartree-Fock mean field, we express the $\hat{H}_0+\hat{H}_{\rm int}$ in the momentum space and decouple all four-fermion operator terms by evaluating all their possible contractions. 
For example, for nearest-neighbor C to B repulsion in $\hat{H}_{\rm int}$, it can be written as
\begin{equation}
\begin{aligned}
\sum_{\langle l, m\rangle} U \hat{C}_l^{\dagger} \hat{B}_m^{\dagger} \hat{B}_m \hat{C}_l & =\frac{U}{N_{s}} \sum_{\mathbf{k}_1 \mathbf{k}_{\mathbf{k}} \mathbf{k}_{\mathbf{3}} \mathbf{k}_4} \sum_{i=2,4,6} \delta_{\mathbf{k}_3+\mathbf{k}_4-\mathbf{k}_1-\mathbf{k}_2,\mathbf{0}}  \\
&\times e^{-i\left(\mathbf{k}_4-\mathbf{k}_2\right) \cdot \mathbf{d}_i}  \hat{C}_{\mathbf{k}_1}^{\dagger} \hat{B}_{\mathbf{k}_2}^{\dagger} \hat{B}_{\mathbf{k}_4} \hat{C}_{\mathbf{k}_3}.
\end{aligned}
\end{equation}
Replacing the four-fermion operator terms with all their possible contractions, this repulsion term is reduced to its Hartree-Fock mean-field form $H^{\rm CB}_{\rm int,HF}$, which reads
\begin{equation}
\begin{aligned}
H^{\rm CB}_{\rm int,HF} & =\frac{3U}{N_s} \sum_{\mathbf{k}\mathbf{k}_1}\langle\hat{C}_{\mathbf{k}_1}^{\dagger} \hat{C}_{\mathbf{k}_1}\rangle \hat{B}_{\mathbf{k}}^{\dagger} \hat{B}_{\mathbf{k}}+\frac{3U}{N_s} \sum_{\mathbf{k}\mathbf{k}_1}\langle\hat{B}_{\mathbf{k}_1}^{\dagger} \hat{B}_{\mathbf{k}_1}\rangle \hat{C}_{\mathbf{k}}^{\dagger} \hat{C}_{\mathbf{k}} \\
& -\frac{U}{N_s} \sum_{\mathbf{k}\mathbf{k}_1} \sum_{i=2,4,6}e^{-i\left(\mathbf{k}_1-\mathbf{k}\right) \cdot \mathbf{d}_i}\langle\hat{C}_{\mathbf{k}_1}^{\dagger} \hat{B}_{\mathbf{k}_1}\rangle \hat{B}_{\mathbf{k}}^{\dagger} \hat{C}_{\mathbf{k}}\\
&-\frac{U}{N_s} \sum_{\mathbf{k}\mathbf{k}_1}\sum_{i=2,4,6}e^{-i\left(\mathbf{k}-\mathbf{k}_1\right) \cdot \mathbf{d}_i}\langle\hat{B}_{\mathbf{k}_1}^{\dagger} \hat{C}_{\mathbf{k}_1}\rangle \hat{C}_{\mathbf{k}}^{\dagger} \hat{B}_{\mathbf{k}}.
\end{aligned}
\end{equation}
Applying this Hartree-Fock mean-field to all repulsion terms in $\hat{H}_{\rm int}$ gives rise to $\hat{H}_{\rm int, HF}$.
The order parameters $\langle\hat{\alpha}_{\mathbf{k}}^\dagger \hat{\beta}_{\mathbf{k}}\rangle$ with $\alpha,\beta=A, B, C$ are ground-state operator expectations of the Hartree-Fock mean-field Hamiltonian $\hat{H}_0+H_{\rm int,HF}$. The solution is got in a self-consistent way.

\begin{figure}
\centering
\includegraphics[width=1\columnwidth]{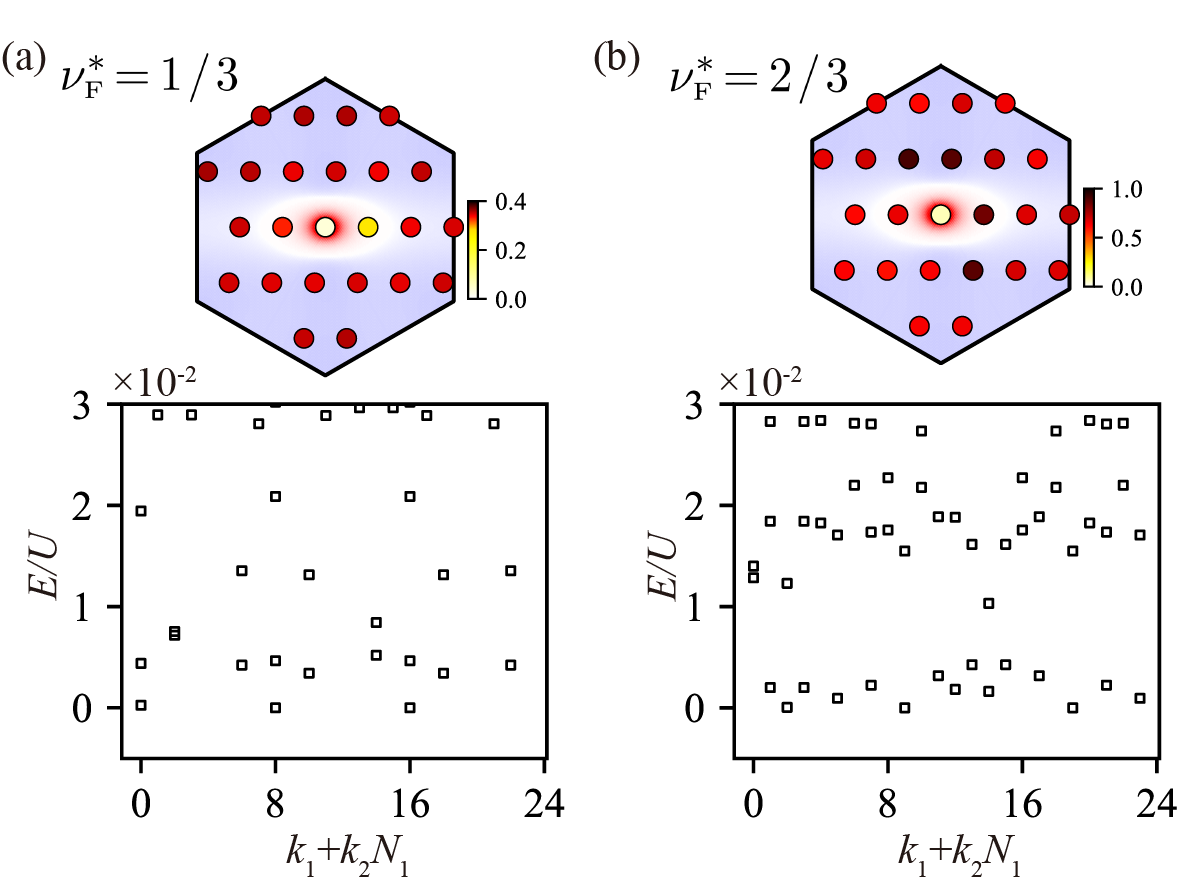}
\caption{\textbf{ED spectra and corresponding hole distributions obtained from calculations on the exact flat band.}
(a) ED spectrum with hole filling of $\nu^{\ast}_{\rm F}=1/3$ (the lower panel) and the averaged hole distribution $\tilde{n}_{\mathbf{k}}$ over the three degenerate ground states (the upper panel). (b)  ED spectrum at $\nu^{\ast}_{\rm F}=2/3$ (the lower panel) and the hole distribution $\tilde{n}_{\mathbf{k}}$ for the lowest state in K=9 sector (the upper panel). The hole filling of $\nu^{\ast}_{\rm F}=1/3$ ($\nu^{\ast}_{\rm F}=2/3$) is equivalent to the electron filling of $\nu_{\rm F}=2/3$ ($\nu_{\rm F}=1/3$). The $4\times6$ $k$-mesh 2 is used, $\eta=0.7$ and $U=0.001$.
}\label{Fig_flatband-1-3}
\end{figure}

\section{Comparison of two-band and one-band projected ED}
\label{APP_Comparison}

\begin{figure}
\centering
\includegraphics[width=1\columnwidth]{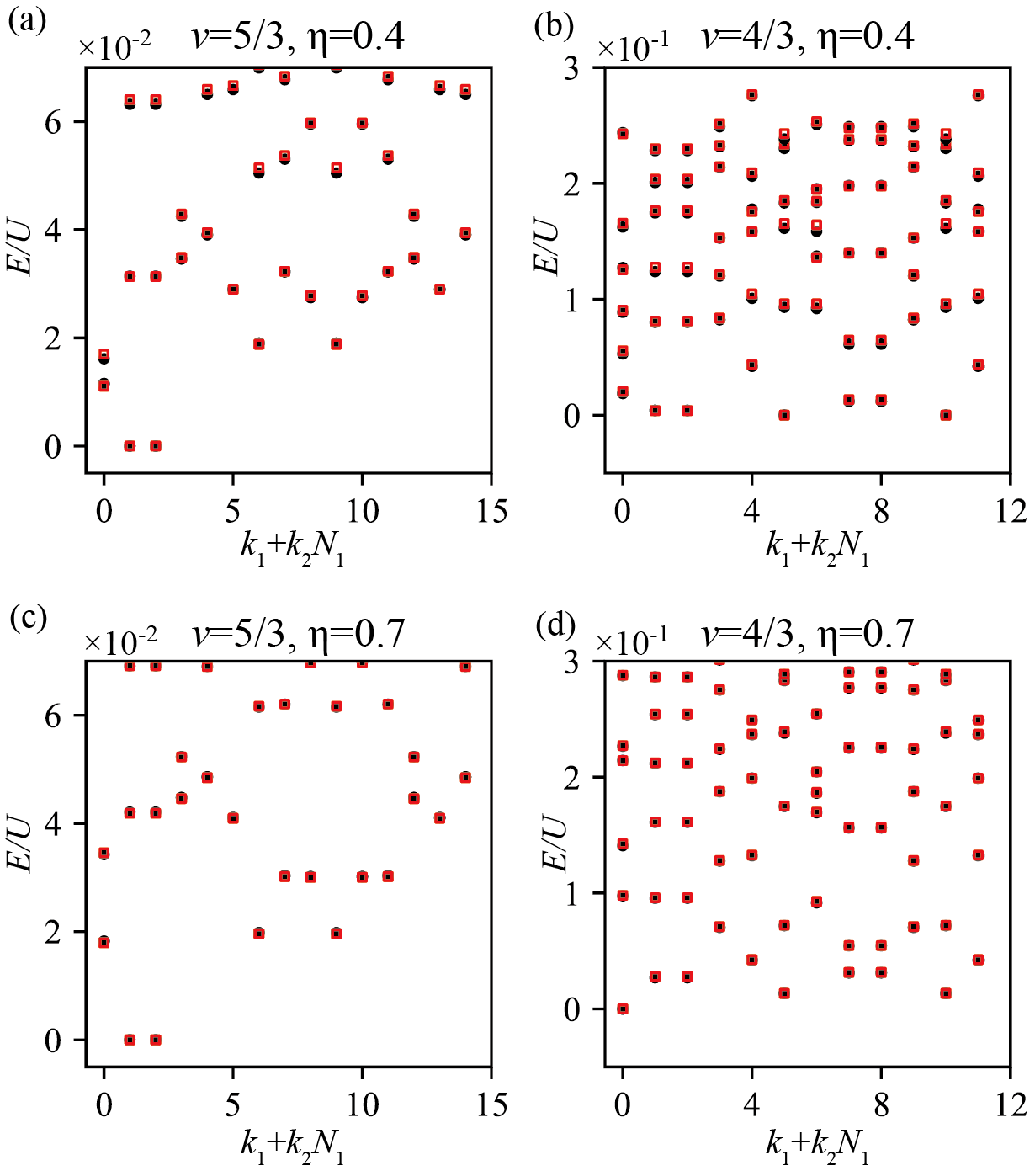}
\caption{(a) Many-body spectra at overall filling $\nu=5/3$ (flat band filling $\nu_\mathrm{F}=2/3$) with $\eta=0.4$ and $U=0.01$, comparing two-band ED (black circles) and one-band projected ED with Hartree-Fock renormalization (red squares), on a $3\times5$ $k$-mesh 1 (the definition of the $k$-mesh 1 is given in \ref{APP_Competition}). (b) Similar comparison for $\nu=4/3$ (flat band filling $\nu_{\rm F}=1/3$), $\eta=0.4$ and $U=0.01$, on a $3\times4$ $k$-mesh 1. (c) Similar comparison for $\nu=5/3$ (flat band filling $\nu_\mathrm{F}=2/3$), $\eta=0.7$ and $U=0.001$, on a $3\times5$ $k$-mesh 1. (d) Similar comparison for $\nu=4/3$ (flat band filling $\nu_\mathrm{F}=1/3$), $\eta=0.7$ and $U=0.001$, on a $3\times4$ $k$-mesh 1.
}\label{Fig_benchmark}
\end{figure}

\begin{figure}
\centering
\includegraphics[width=1\columnwidth]{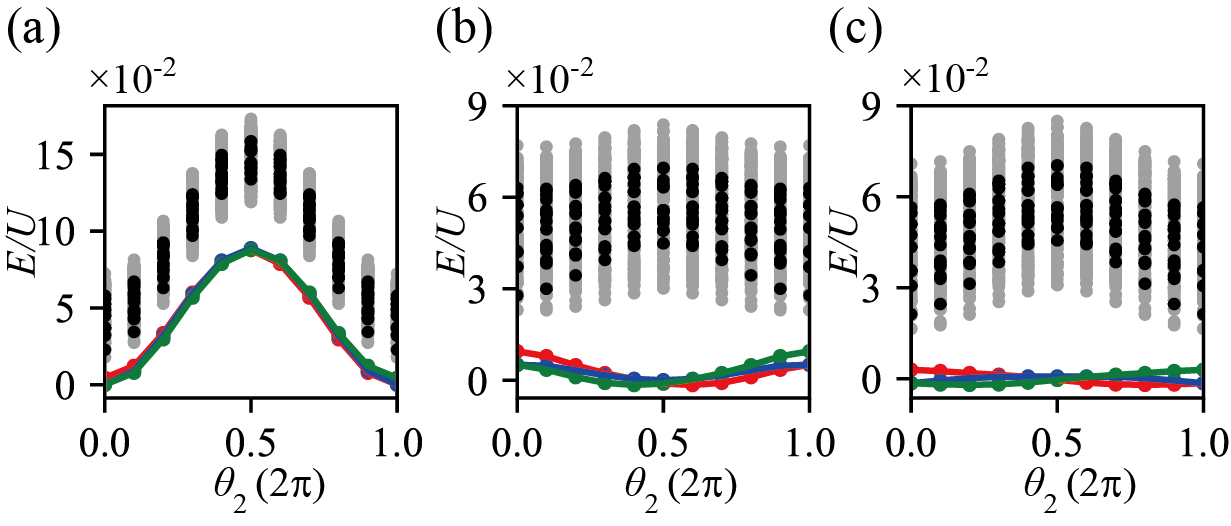}
\caption{(a)-(c) Spectral flows as a function of phase twist angle $\theta_2$ without  background energy subtraction (a) and with background energy subtraction, estimated by computing the mean value of the diagonal elements of the Hamiltonian (b) and averaging the energies of the three degenerate ground states (c)  (c.f. \ref{APP_Background}), based on a $4\times6$ $k$-mesh 1. The convention for $\theta_2$ is provided in \ref{APP_Competition}. Red, blue, and green scatters indicate the lowest states in the three expected ground-state momentum sectors for a FCI on a torus, black ones denote excited states within these sectors, and gray ones label all other states. Only the 6 lowest states for each momentum sector are shown. Parameters: $\eta=0.4$ and $U=0.01$.
}\label{Fig_back}
\end{figure}

\begin{figure}[!htp]
\centering
\includegraphics[width=1\columnwidth]{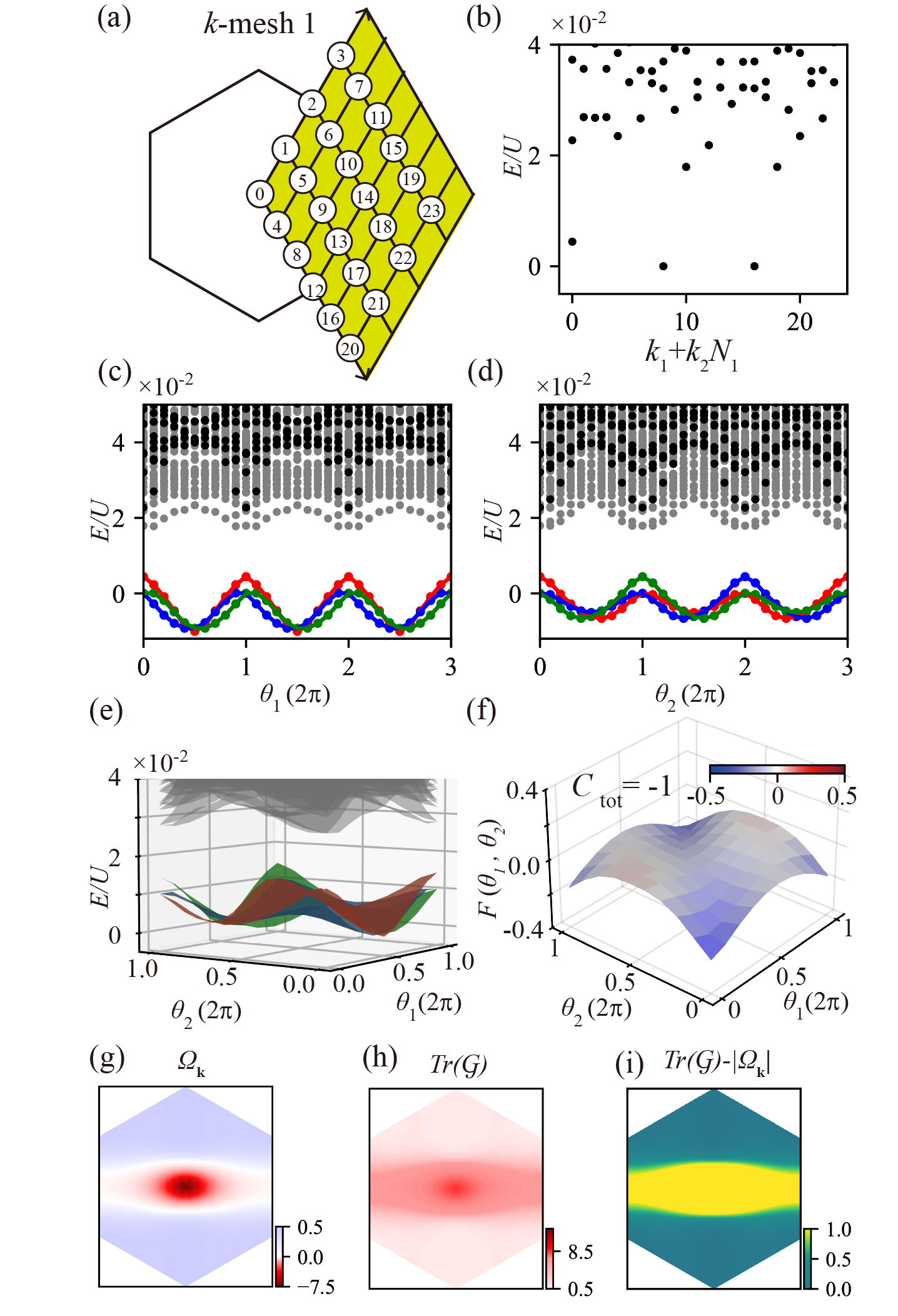}
\caption{(a) Illustration of the $k$-mesh 1 of a size of $4\times6$, with all momenta indexed. (b) ED spectrum based on $k$-mesh 1 for flat band filling $\nu_{\rm F}=2/3$, where the total many-body Chern number is -1. (c) Spectral flow as a function of phase twist angle $\theta_1$, at $\theta_2=0$. (d) Spectral flow as a function of phase twist angle $\theta_2$, at $\theta_1=0$. (e) Spectral flow in the two-dimensional twisted boundary condition space.
(f) Distribution of many-body Berry curvature $F(\theta_1,\theta_2)$. (g) Berry curvature $\Omega_{\mathbf{k}}$. (h) Trace of quantum metric tensor $\mathcal{G}$. (i) ${\rm Tr} \mathcal{G}-|\Omega_{\mathbf{k}}|$.
The convention for $\theta_1$ and $\theta_2$ is provided in \ref{APP_Competition}.
In spectral flow plots,
we have subtracted a background energy from the spectrum at every phase twist angle $\theta$, for a better visualization of the spectral gap as a function of $\theta$.
Red, blue, and green scatters indicate the lowest states in the three expected ground-state momentum sectors for a FCI on a torus, black ones denote excited states within these sectors, and gray ones label all other states.
Parameters: $\eta=0.4$ and $U=0.01$, and $\nu_{\rm F}=2/3$.
}\label{Fig_ED46}
\end{figure}

As a further justification of the single band ED in the isolated band limit, we compare the ED calculations of two approaches: (1) fully encompassing the two lowest bands, and (2) with the single band taking into account the Hartree-Fock renormalization. The comparisons are carried out on $3\times4$ and $3\times5$ unit-cell clusters.
The ED energy spectra from these two approaches are consistent for both flat band filling factors $\nu_\mathrm{F}=1/3$ and $2/3$, proving the validity of the single band ED (Fig.~\ref{Fig_benchmark}). This consistency can be expected as we focus on the isolated band limit, i.e., with $\eta$ values within the gray shaded areas in  Figs. \ref{Fig_HF}(e) and \ref{Fig_HF}(f), where interaction renormalization leaves the flat band quantum geometry unchanged. 
In this regime, the particles filling the lower band are essentially frozen and their influence manifests as modifications to the diagonal energies of particles filling the flat band.
This diagonal modification is accounted by the self-consistent Hartree-Fock renormalization of the flat band dispersion [c.f. Fig.~\ref{Fig_BC}(e)].

\begin{figure}[!htp]
\centering
\includegraphics[width=1\columnwidth]{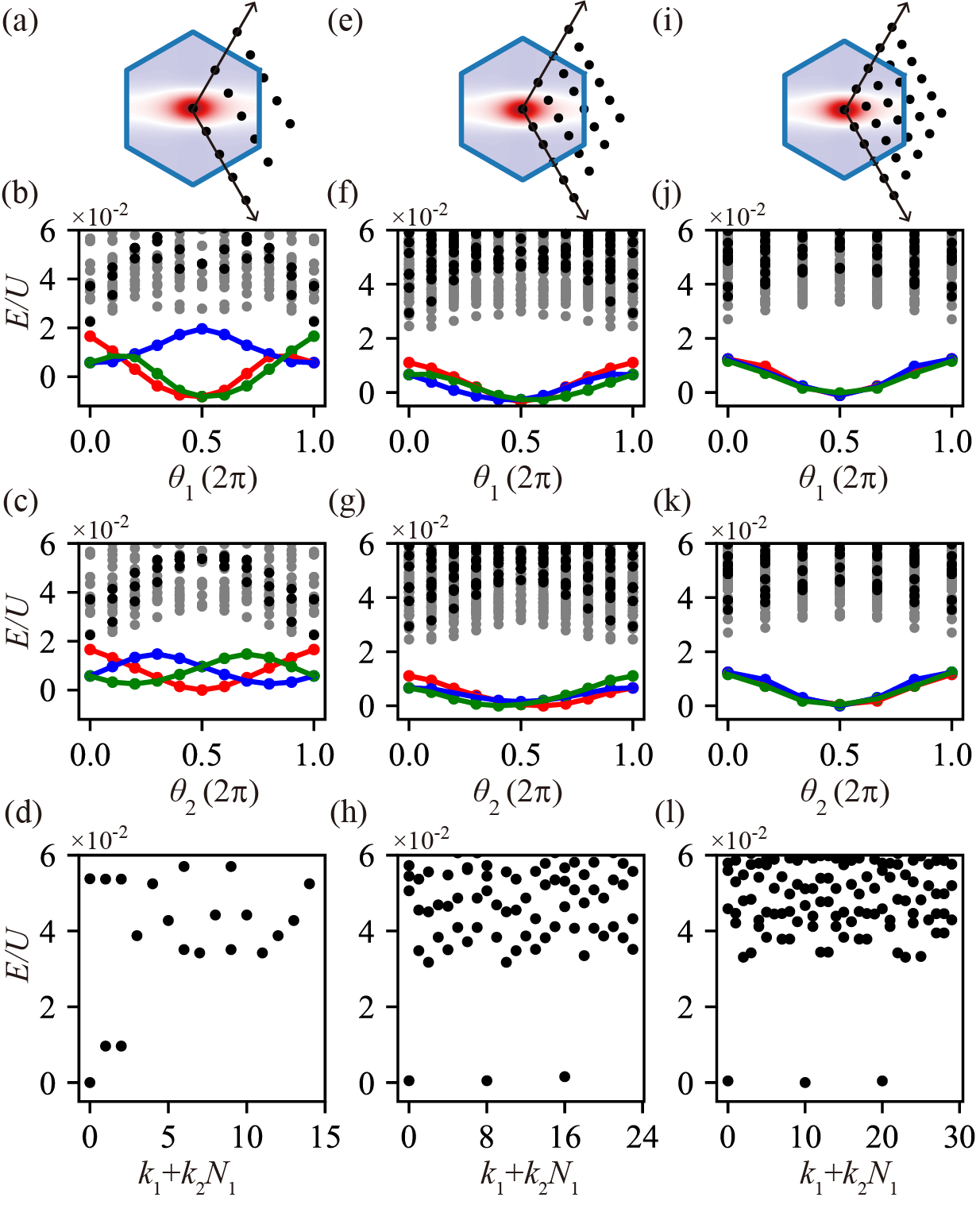}
\caption{(a)-(d) ED spectra obtained with a $3\times 5$ $k$-mesh 1.
(a) Schematic of the $k$-mesh, with BZ color coded with Berry curvature distribution as an indication of anisotropy.
(b) Spectral flow as a function of phase twist angle $\theta_1$, at $\theta_2=0$.
(c) Spectral flow as a function of phase twist angle $\theta_2$, at $\theta_1=0$.
(d) ED spectrum at $(\theta_1,\theta_2)=(0,\pi)$. (e)-(h) Similar plots, but for a $4\times 6$ $k$-mesh 1. (i)-(l) Similar plots for a $5\times 6$ $k$-mesh 1.
The convention for $\theta_1$ and $\theta_2$ is provided in \ref{APP_Competition}.
In spectral flow plots,
red, blue, and green scatters indicate the lowest states in the three expected ground-state momentum sectors for a FCI on a torus, black ones denote excited states within these sectors, and gray ones label all other states.
Parameters: $\eta=0.4$ and $U=0.01$ (c.f. Fig.~\ref{Fig_BC} for the Hartree-Fock band character), and $\nu_{\rm F}=2/3$.
}\label{Fig_56-1}
\end{figure}

\begin{figure}
\centering
\includegraphics[width=1\columnwidth]{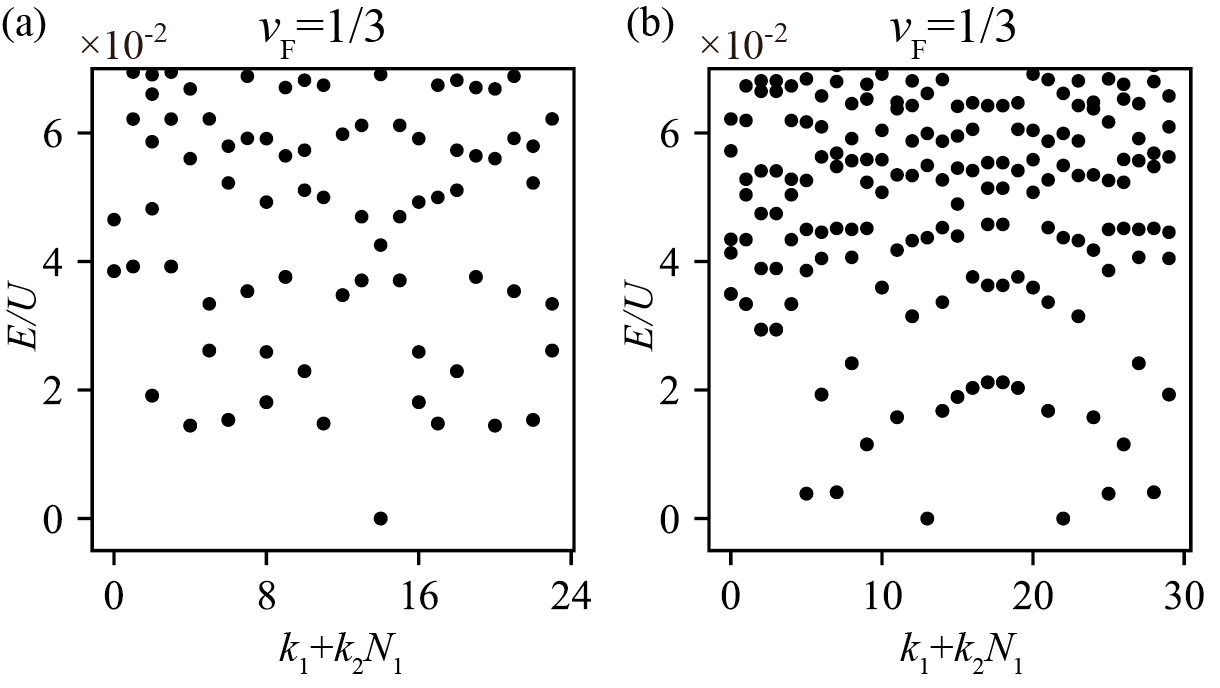}
\caption{Many-body spectrum at flat band filling $\nu_\mathrm{F}=1/3$ (overall filling $\nu=4/3$), from one-band projected ED calculation taking into account the Hartree-Fock renormalization (c.f. Fig.~\ref{Fig_BC}). (a)  With a $4\times6$ $k$-mesh 1. (b) With a $5\times6$ $k$-mesh 1.  Parameters: $\eta=0.4$ and $U=0.01$. In contrast to the $\nu_{\rm F}=2/3$ filling, the ground state is topologically trivial at $\nu_{\rm F}=1/3$ filling.
}\label{Fig_nu1-3}
\end{figure}

\begin{figure*}[!htp]
\centering
\includegraphics[width=0.7\textwidth]{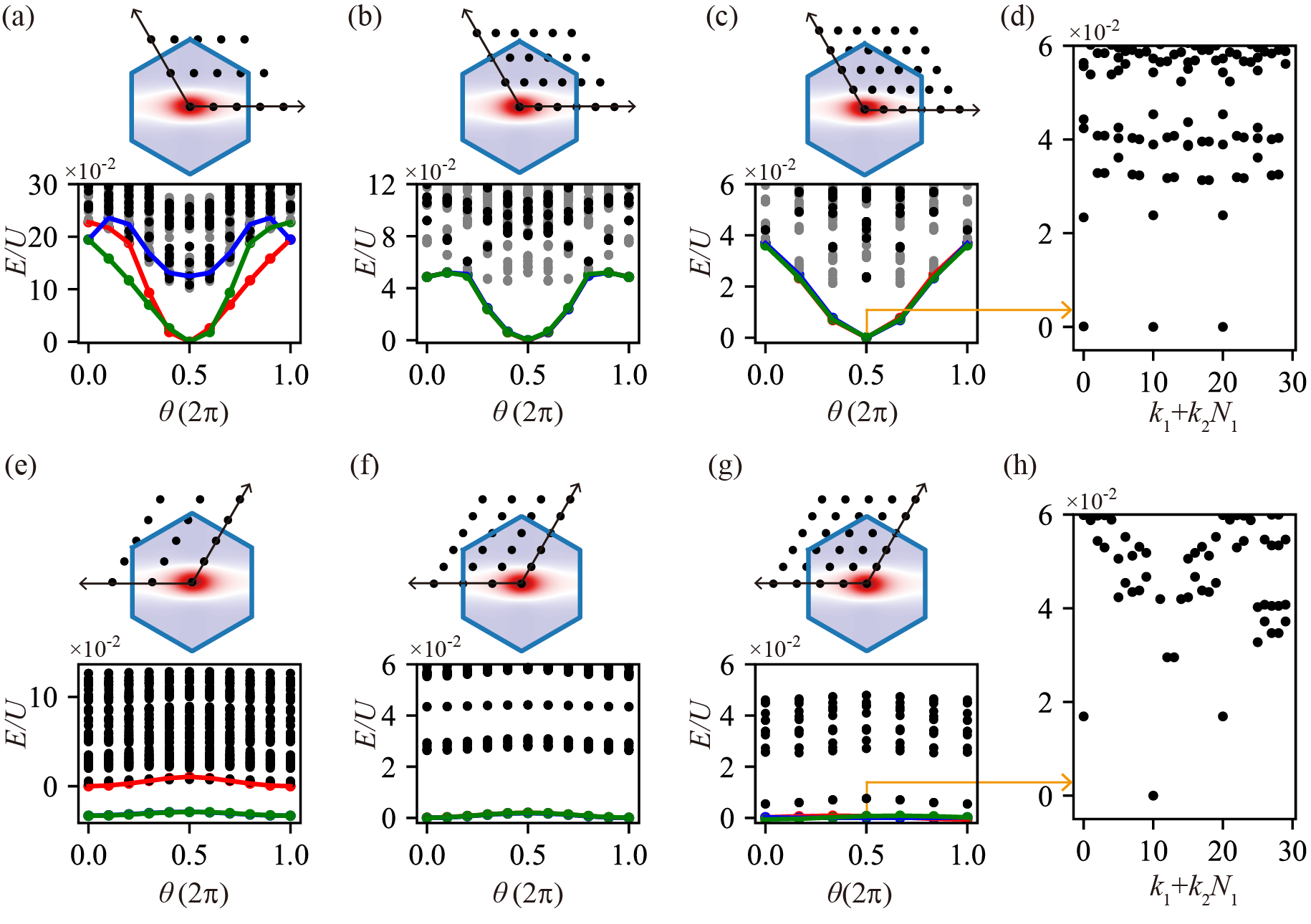}
\caption{(a)-(d) ED spectra calculated on the $k$-mesh 2 with different size: (a) $3\times 5$, (b) $4\times 6$, and (c) $5\times 6$. The upper panels illustrate the schematics of the  $k$-meshes. The lower panels display the spectral flows along the upward-tilting axes, whereas spectral flows undergo minimal changes in the horizontal direction. 
(d) ED many-body spectrum for the  $5\times 6$ $k$-mesh 2 with $\theta = \pi$ [c.f. orange right-angled arrow in (c)].
(e)-(h) Similar plots, but for the $k$-mesh 3.
In spectral flow plots, red, blue, and green scatters indicate the lowest states in the three expected ground-state momentum sectors for a FCI on a torus, black ones denote excited states within these sectors, and gray ones label all other states.
Parameters: $\eta=0.4$ and $U=0.01$ (c.f. Fig.~\ref{Fig_BC} for the Hartree-Fock band character), and $\nu_{\rm F}=2/3$.
}\label{Fig_56-2}
\end{figure*}

\begin{figure}[!htp]
\centering
\includegraphics[width=0.9\columnwidth]{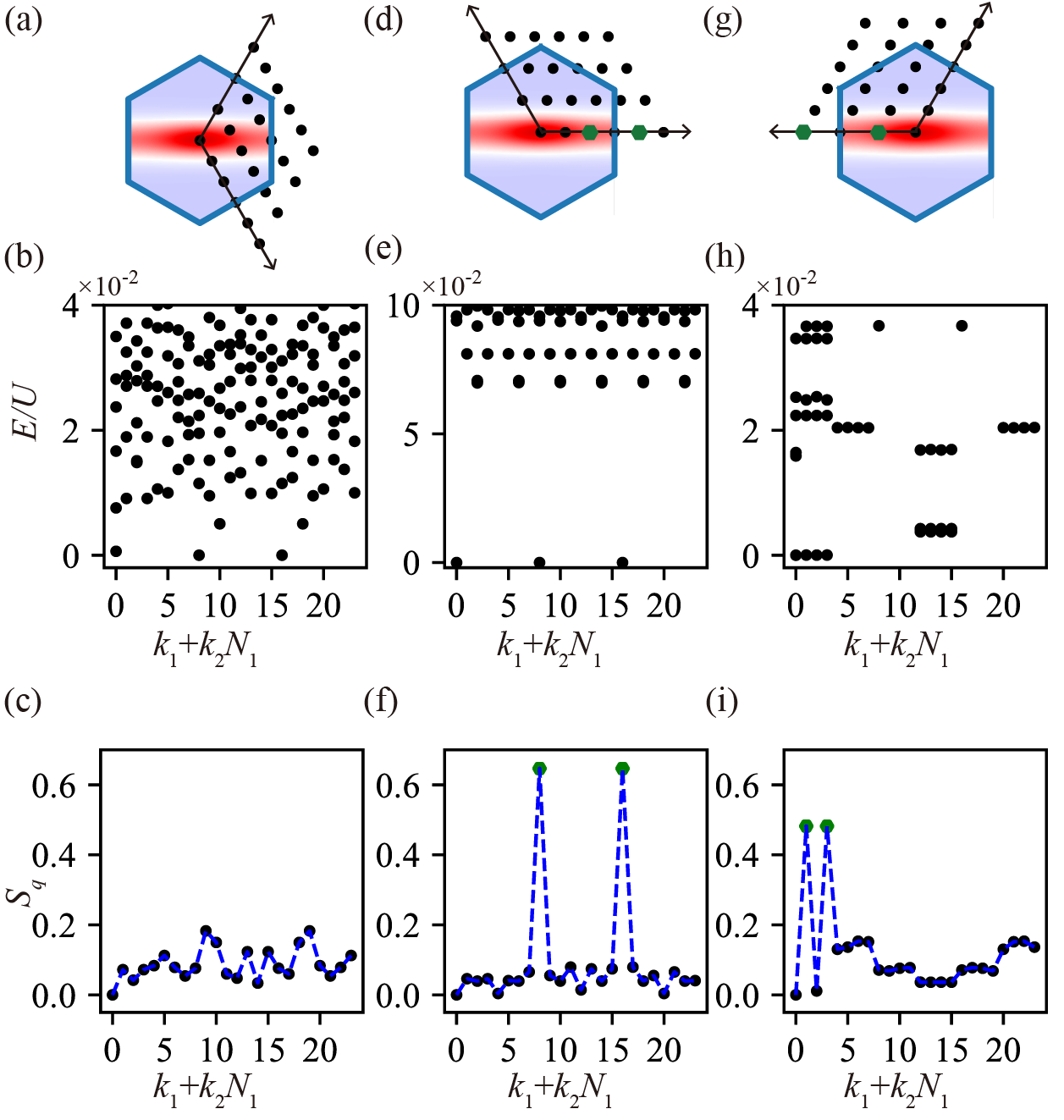}
\caption{(a)-(c) ED calculations on a $4\times6$ $k$-mesh 1. (a) Illustration of the $4\times6$ $k$-mesh 1. (b) ED spectrum. (c) Structure factor for the lowest state at K=0 momentum sector. (d)-(f) Similar plots, but for a $4\times6$ $k$-mesh 2. (g)-(i) Similar plots for a $4\times6$ $k$-mesh 3.
Parameters: $\nu_{\rm F}=2/3$, $\eta=0.1$ and $U=0.01$.
}\label{Fig_phase_1}
\end{figure}

\begin{figure}
\centering
\includegraphics[width=1\columnwidth]{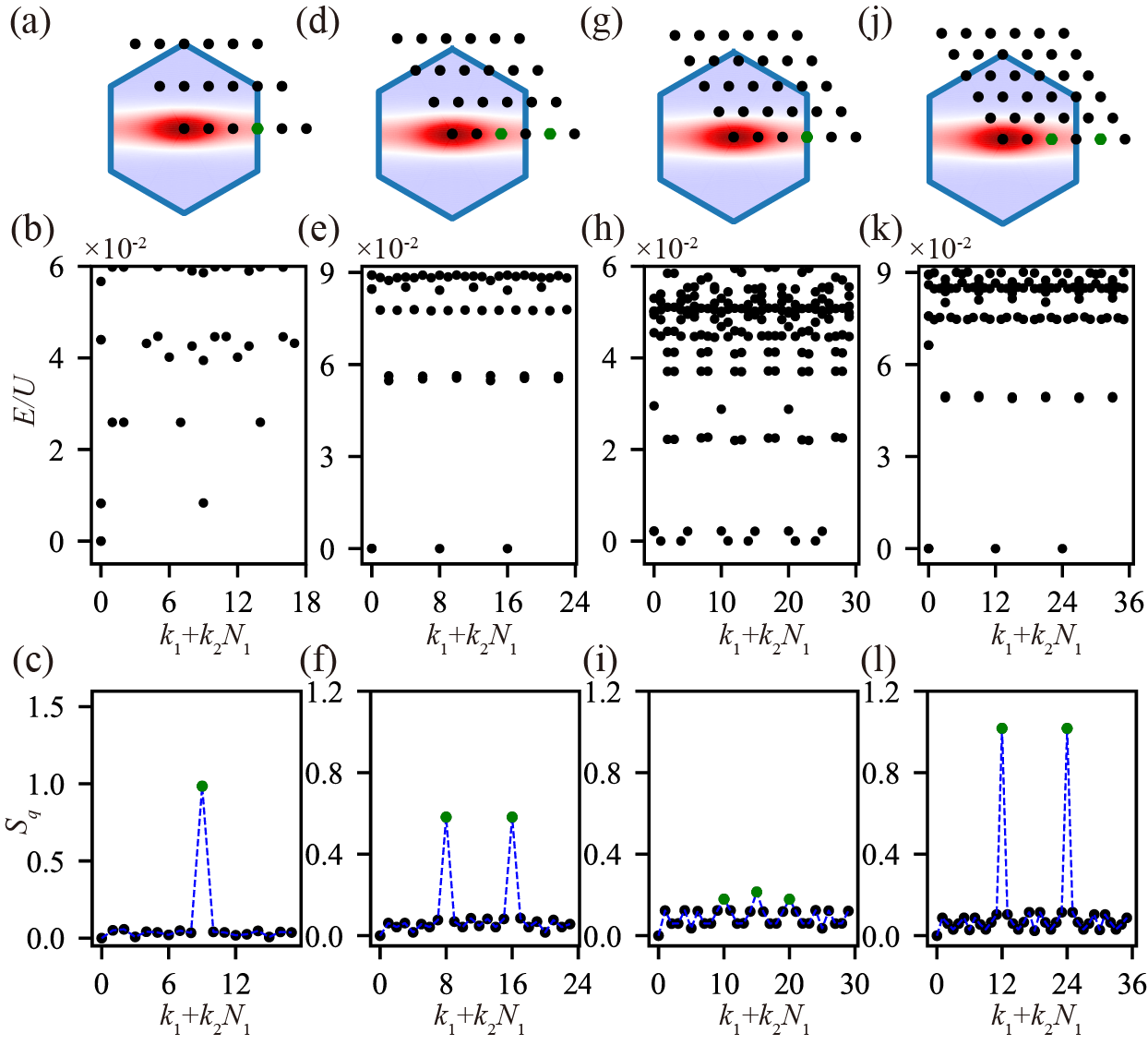}
\caption{(a)-(c) ED calculations on a $3\times6$ $k$-mesh 2. (a) Illustration of the $3\times6$ $k$-mesh 2. (b) ED spectrum. (c) Structure factor averaged over a few low-energy states, where the peak's $\mathbf{q}$ vector highlighted by the green hexagon in (a).  (d)-(f) Similar plots for a $4\times6$ $k$-mesh 2. (g)-(i) Similar plots for a $5\times6$ $k$-mesh 2.  (j)-(l) Similar plots for a $6\times6$ $k$-mesh 2.
Parameters: $\nu_{\rm F}=2/3$, $\eta=0.2$ and $U=0.01$.
}
\label{Fig_cdw_comparison}
\end{figure}

\section{Hartree-Fock renormalized flat band dispersion and emergent kinetic energy}
\label{APP_HF_renormalized}
The Hartree-Fock renormalized flat band dispersion can alternatively be obtained by evaluating the expectation values of terms involving $d_{1 \mathbf{k}}$ in the interaction Hamiltonian $\hat{H}_{\rm int}$ with respect to the vacuum state, defined as $| \rm Vac\rangle=\prod_{\mathbf{k}} d_{1 \mathbf{k}}^{\dagger}|0\rangle$, where $d_{n, \mathbf{k}}(d_{n, \mathbf{k}}^{\dagger})$ is the annihilation (creation) operator for the eigenstates of the fluxed dice lattice model. These operators have a transformational relationship with the original operators,
$$
\alpha_k^{\dagger}=\sum_n u_{\alpha, n}(\mathbf{k}) d_{n, \mathbf{k}}^{\dagger},
$$
where $\alpha= A,B,C$ and $u_{\alpha, n}(\mathbf{k})$ is the periodic part of the Bloch wave function. For example, the terms in Hartree-Fock renormalized flat band dispersion contributed by the B to C repulsion reads
\begin{small}
\begin{equation}
-\frac{U}{N_s} \sum_{\mathbf{k}^{\prime}} \sum_{n_1 n_2=1,2} \sum_{i=2,4,6} e^{-i\left(\mathbf{k}^{\prime}-\mathbf{k}\right) \cdot \mathbf{d}_i} u_{C, n_2}\left(\mathbf{k}^{\prime}\right)u_{B, n_1}(\mathbf{k}) u_{B, n_2}^*\left(\mathbf{k}^{\prime}\right) u_{C, n_1}^*(\mathbf{k}),
\end{equation}
\end{small}
where $n_1\neq n_2$. It is noted that the dispersion from Hartree-Fock renormalization can be expressed in terms of the quantum metric of the filled dispersive band, which is identical to that of the flat band.

The Hamiltonian $H^{\rm proj}$ can also be analyzed within the hole representation. Under particle-hole transformation ($\hat{d}_{2 \mathbf{k}}^{\dagger} \rightarrow \hat{h}_{2,-\mathbf{k}}, \hat{d}_{2 \mathbf{k}} \rightarrow \hat{h}_{2,-\mathbf{k}}^{\dagger}$), an emergent kinetic energy arises that precisely cancels the Hartree-Fock renormalized flat band dispersion $\varepsilon_{\mathbf{k}}$ in the two-band model. The intrinsic link between Hartree-Fock energy and hole kinetic energy has been previously reported ~\cite{Abouelkomsan2023_metric_induce}.
As a result, the particle-hole transformation yields a Hamiltonian $\tilde{H}^{\rm proj}$ whose single-particle band is flat. The emergence of such an exact flat band in the hole representation becomes evident if one initially transforms the electron picture to hole picture in the two-band model. 

\section{Background energy subtraction in the spectral flow}
\label{APP_Background}
In spectral flow plots presented in the main text, at every phase twist angle $\theta$, we have subtracted a background energy, which depends on $\theta$, for the energy spectrum. Initially, the background energy is estimated by computing the mean value of the diagonal elements of the Hamiltonian for each $\theta$, a method employed in the spectral flow plots presented in \ref{APP_Competition}. However, it is more convenient to estimate the background energy directly by averaging the energies of the three degenerate ground states; this approach is utilized in the spectral flow plots in the main text and \ref{App_Supporting}.
The anisotropy of the model  results in a considerable change in the background energy with $\theta$. A comparison of the spectral flow without and with background energy (estimated using both methods) subtraction is shown in Fig. \ref{Fig_back}. The background energy subtraction does not affect the spectral gap as a function of $\theta$, but rather allows a better visualization of this dependence.

\section{Competition of FCI and CDW at $\nu_\mathrm{F}=2/3$ with stronger anisotropy ($\eta$ = 0.4) under the finite sizes}
\label{APP_Competition}

In the main text, we have clearly addressed the CDW order at small $\eta$, with a tripled inter-chain period. This might suggest the finite-size effect if the small ED clusters are not compatible with the order especially those without the momenta of the CDW wave vector.
This motivates us to reexamine the results from smaller clusters  such as $3\times5$, $4\times6$ and $5\times6$,  as reported in the previous version of this paper \cite{lin2025fractional}. In these situations, FCI states can form at $\eta=0.4$, where the anisotropy is more pronounced compared to $\eta=0.7$ adopted in the calculations for the $6\times6$ cluster.
In contrast to the $6 \times 6$ cluster now adopted, some of these small clusters are not compatible with the CDW order and can therefore extend the FCI states toward stronger anisotropy.
In this section, we show these ED results in a context for discussing how FCI and CDW compete at stronger anisotropy  ($\eta=0.4$) in these smaller clusters. In this section, unless explicitly stated otherwise, we consider the filling factor $\nu_{\rm F}=2/3$.

 When ED calculations are conducted based on the clusters of size $3\times5$, $4\times6$ and $5\times6$,
in addition to the mesh identified in Figs.~\ref{Fig_ED6x6}(a) and ~\ref{Fig_ED6x6}(f) (previously referred to as $k$-mesh 2), one can also rotate the reciprocal primitive vectors by 60$^{\circ}$  clockwise and counterclockwise, leading to two different meshes, namely $k$-mesh 1 and $k$-mesh 3. For the sake of simplicity of the symbols, the flux along $\theta_1$ in $k$-mesh 1 ($k$-mesh 3) is to be interpreted as along the reciprocal lattice vector obtained by rotating  $\mathbf{b}_1$ clockwise (counterclockwise) by 60$^{\circ}$, and the flux along $\theta_2$ as that obtained by rotating $\mathbf{b}_2$ similarly.
ED calculations on these clusters using the three different $k$-meshes yield distinct results, which manifests the presence of anisotropy.

 For the $4\times6$ $k$-mesh 1, robust FCI states already form at $\eta=0.4$, as evidenced by the three-fold degenerate ground states [Figs. \ref{Fig_ED46}(a) and \ref{Fig_ED46}(b)], the persistent gap in the spectral flow [Figs. \ref{Fig_ED46}(c)-\ref{Fig_ED46}(e)], and the total Chern number of -1 [Fig. \ref{Fig_ED46}(f)]. ED results of clusters with $3 \times 5$, $4\times6$ and $5\times 6$ unit-cells are compared in Fig. \ref{Fig_56-1}. It is evident that the degeneracy of three FCI states is significantly improved in larger clusters, and the excitation gap increases, suggesting that expanding the scale of the cluster strengthens the robustness of the FCI states. In contrast to the $\nu_{\rm F}=2/3$, the ground state is topologically trivial at $\nu_{\rm F}=1/3$ filling [Fig. \ref{Fig_nu1-3}].

 For $k$-mesh 2 of sizes $3\times 5$ and $4 \times 6$, the ground states overlap with the excited states in the spectral flow upon flux insertion [Figs. \ref{Fig_56-2}(a) and \ref{Fig_56-2}(b)]. The many-body Chern numbers are also found to be zero for ED calculations on these two configurations. Such an overlap disappears in the $5 \times 6$ cluster [Fig. \ref{Fig_56-2}(c)], and a nearly perfect degeneracy of the ground state in ED many-body spectrum for $(\theta_1,\theta_2)=0,\pi/2)$ is observed [Fig. \ref{Fig_56-2}(d)]. The total many-body Chern numbers for these ground states are precisely -1.

For $k$-mesh 3, the overlap between ground states and excited states in the spectral flow is more significant for smaller-size cluster, as compared to larger-size clusters [Figs. \ref{Fig_56-2}(e)-\ref{Fig_56-2}(h)].
Nontrivial total many-body Chern number of -1 are already obtained for the $4 \times 6$ cluster. Similarly, for the $5 \times 6$ cluster, the total many-body Chern numbers is also -1. In both cases, the ground states still have some spectral overlap with the excited states, but they are from different momentum sectors.

In these smaller clusters, we find that the CDW order identified in the main text is well captured only by the $4\times6$ $k$-mesh 2. This is evidenced by the ED spectra at  $\eta=0.1$ for the three $k$-meshes,  shown in Figs. \ref{Fig_phase_1}(b), \ref{Fig_phase_1}(e) and \ref{Fig_phase_1}(h), as well as by the corresponding structure factors [Figs. \ref{Fig_phase_1}(c), \ref{Fig_phase_1}(f) and \ref{Fig_phase_1}(i)]. Here, $\eta=0.1$ corresponding to even stronger anisotropy is chosen to suppress FCI states in the $k$-mesh 1, thereby enabling a focused study of the candidate CDW states across all $k$-meshes.
While both $k$-mesh 1 and $k$-mesh 2 exhibit a three-fold degenerate ground state, $k$-mesh 3 does not.
Notably, only $k$-mesh 2 displays distinct peaks in $S(\mathbf{q})$ at $\pm1/3\mathbf{b}_2$, highlighting the presence of characteristic CDW ordering.

Within the $k$-mesh 2, even if the number of unit cells along the inter-chain direction is compatible with the tripled period, there might be some even-odd effect at small sizes [Fig. \ref{Fig_cdw_comparison}].
More specifically, in the finite-size ED, the CDW order is only well established when the number of unit cells along the chain is even, which refers to the $4\times6$ and $6\times6$ cases.
When the number of unit cells along the chain is odd, for example at $3\times6$, we find the Bragg peak supports a different doubled inter-chain period (also supported by the low-lying state at $\mathbf{b}_2$).
However, at $5\times6$, the corresponding Bragg peak of the doubled inter-chain period gets much less pronounced, suggesting that this order might be a finite-size artifact, not favored in the thermodynamic limit. The spectra of the $5\times6$ also becomes very complicated, possibly due to competitions among different states such as the gradually vanishing $\mathbf{b}_2/2$ CDW, and the expected $\mathbf{b}_2/3$ CDW.
Note that, this even-odd effect does not matter in the thermodynamic limit, which is supported by our iDMRG results in the main text, where the $\mathbf{b}_2/3$ CDW is robust when the number of unit cells along the disconnected chains is infinite.

Therefore, we have found that the suppression of the CDW order promotes the formation of the FCI state, as discussed below. For $k$-mesh 1, neither of the primitive vectors lies along the inter-chain direction, which strongly suppresses the formation of CDW states. This explain the emergence of the robust FCI states at smaller $\eta$ value compared to those in $6 \times 6$ clusters. In contrast, when $k$-mesh 2 of size $4 \times 6$ is employed—a configuration that is compatible with the identified CDW order—the many-body Chern number vanishes. The zero Chern number is a consequence of an anticrossing observed in the spectral flow, indicating the suppression of FCI states under these conditions [Fig.~\ref{Fig_56-2}(b)]. When the system size is increased to $5 \times 6$ for this mesh, FCI states reemerge. This may be attributed to the even-odd effect in this mesh that is unfavorable for the expected inter-chain CDW, even if the other axis is oriented along the inter-chain direction. For $k$-mesh 3, one of its axes is also aligned with the inter-chain direction. As a result, there is a stronger overlap between the ground and excited states, leading to poorer ground state degeneracy, despite the many-body Chern number of -1.

\section{Supporting results of the FCI from different k-meshes at $\eta=0.7,\ U=0.001$}
\label{App_Supporting}

\begin{figure}
\centering
\includegraphics[width=1\columnwidth]{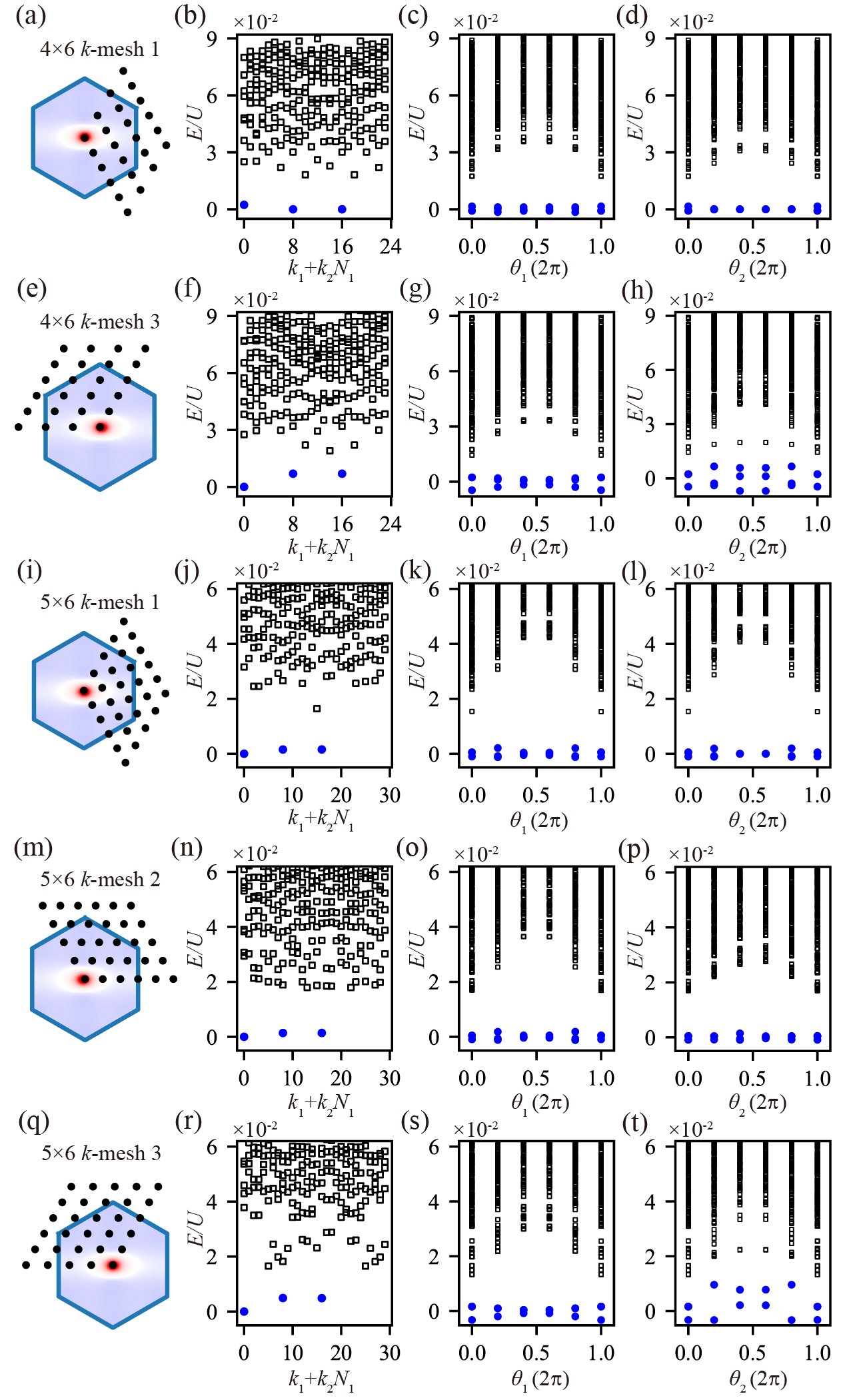}
\caption{(a)-(d) ED spectra obtained with a $4\times 6$ $k$-mesh 1.
(a) Schematic of the $k$-mesh, with BZ color coded with Berry curvature distribution.
(b) Spectral flow as a function of phase twist angle $\theta_1$, at $\theta_2=0$.
(c) Spectral flow as a function of phase twist angle $\theta_2$, at $\theta_1=0$.
(e)-(h) Corresponding results for a  $4\times 6$ $k$-mesh 3.
(i)-(l) Corresponding results for a  $5\times 6$ $k$-mesh 1.
(m)-(p) Corresponding results for a  $5\times 6$ $k$-mesh 2.
(q)-(t) Corresponding results for a  $5\times 6$ $k$-mesh 3.
The convention for $\theta_1$ and $\theta_2$ is provided in \ref{APP_Competition}.
Parameters: $\eta=0.7$ and $U=0.001$ (c.f. Fig.~\ref{Fig_BC} for the Hartree-Fock band character), and $\nu_{\rm F}=2/3$.
}\label{Fig_all_size}
\end{figure}

We emphasize that the FCI states remain robust across system sizes of  $4\times6$, $5\times6$ and $6 \times 6$ in the $k$-mesh 1, 2 and 3 (refer to \ref{APP_Competition} for the definition of the $k$-meshes). Specifically, for these systems, the ground state with the expected degeneracy for a FCI on a torus are consistently observed [Fig. \ref{Fig_ED6x6} and Fig. \ref{Fig_all_size}], and the energy gap separating the ground states from the excited states persists throughout the spectral flow.

As mentioned earlier, at finite sizes, when the cluster/mesh is incompatible with the inter-chain CDW with tripled unit cell, the FCI might be more favored. For example, for k-mesh 2, even when the CDW momenta are included, due to the even-odd effect at finite size as discussed, the CDW is suppressed at the $5\times6$ size with odd number of unit cells along the disconnected chain, and the spectral gap might be even slightly larger than that of the $6\times6$ cluster.
However, by comparing the clusters with even (or with odd) numbers of unit cells along the chain separately (we also considered the $3\times6$ size while not plotted here), the overall trend of the spectral gap of this FCI (from isolated trivial band) is that it increases with system size.


%

\end{document}